\documentclass[twocolumn,pra,showpacs,preprintnumbers,amsmath,amssymb]{revtex4}
\usepackage{graphicx}
\usepackage{dcolumn}
\usepackage{bm}

\begin{document}

\title{Dipolar Fermi gases in anisotropic traps}

\author{Aristeu R. P. Lima}
\email{lima@physik.fu-berlin.de}
\affiliation{Institut f\"{u}r Theoretische Physik, Freie Universit\"{a}t Berlin, Arnimallee 14, 14195 Berlin, Germany}
\author{Axel Pelster}
\email{axel.pelster@fu-berlin.de}
\affiliation{Fachbereich Physik, Universit\"{a}t Duisburg-Essen, Lotharstrasse 1, 47048 Duisburg, Germany}
\affiliation{Institut f\"ur Physik und Astronomie, Potsdam Universit\"{a}t, Karl-Liebknecht-Str.~24, 14476 Potsdam, Germany}

\date{\today}
\begin{abstract}
The quest for quantum degenerate Fermi gases interacting through the anisotropic and long-range dipole-dipole interaction is an
exciting and fast developing branch within the cold-atoms research program. Recent experimental progress in trapping, cooling, and controlling polar molecules with large electric dipole moments has, therefore, motivated much theoretical effort. 
In a recent letter, we have briefly discussed the application of a variational time-dependent Hartree-Fock approach to study
 theoretically both the static and the dynamic properties of such a system in a cylinder-symmetric harmonic trap. We focused on 
the hydrodynamic regime, where collisions assure the equilibrium locally. Here, we present a detailed theory, extended 
to encompass the general case of a harmonic trap geometry without any symmetry. After deriving the equations of motion for 
the gas, we explore their static solutions to investigate key properties like the aspect ratios in both real and 
momentum space as well as the stability diagram. We find that, despite the lack of symmetry of the trap, the momentum 
distribution remains cylinder symmetric. The equations of motion are then used to study the low-lying
hydrodynamic excitations, where, apart from the quadrupole and monopole modes, also the radial quadrupole mode is
investigated. Furthermore, we study the time-of-flight dynamics as it represents an important diagnostic tool for quantum gases. We find that the real-space aspect ratios are inverted during the expansion, while the one in momentum space becomes asymptotically unity. In addition, anisotropic features of the dipole-dipole interaction are discussed in detail. These results could be 
particularly useful for future investigations of strongly dipolar heteronuclear polar molecules deep in the quantum 
degenerate regime.
\end{abstract}

\maketitle
\section{Introduction}
Since the achievement of Bose-Einstein condensation (BEC) with a sample of $^{52}$Cr-atoms \cite{PhysRevLett.94.160401} the interest in dipolar quantum gases has strongly increased \cite{1367-2630-11-5-055009,citeulike:4464283}. Recently, the progress towards quantum degenerate polar molecules has pushed this interest even further because they possess electric dipole moments of the order of one Debye and, therefore, are potential candidates to make dipolar effects more accessible to experiments \cite{K.-K.Ni10102008,arXiv:0811.4618,arXiv:0908.3931,citeulike:6565167}.

Concerning dipolar bosonic particles, the field is relatively well understood and has seen a robust development with a remarkable quantitative agreement between experiment and theory. Starting point was the construction of a pseudopotential by Yi and You \cite{PhysRevA.61.041604}. In the Thomas-Fermi regime, where the kinetic energy can be neglected in comparison with the interaction energy, exact solutions of the Gross-Pitaevskii equation have been found for axially symmetric harmonic traps \cite{PhysRevLett.92.250401,eberlein:033618}. Further generalizations to triaxially anisotropic traps even provided the first clear experimental signature of the dipole-dipole interaction (DDI) in the data for the expansion dynamics \cite{PhysRevLett.95.150406,giovanazzi:013621}. In the meantime, collisional control of chromium has been fully demonstrated by using Fesh\-bach resonances to increase the relative importance of the DDI with respect to the contact interaction. As a result, strong dipolar effects have been observed in BECs like the suppression of the characteristic inversion of the aspect ratio during the expansion \cite{strong-pfau}. In addition, the trap configuration could be manipulated to stabilize a purely dipolar BEC \cite{stabilization-pfau} and a new type of 'Bose-nova' experiment beautifully revealed a d-wave symmetry in a dipolar BEC \cite{d-wave-pfau}. Besides that, the anisotropic nature of the DDI is predicted to shift the Bose-Einstein condensation temperature in a characteristic way \cite{glaum:080407,glaum:023604}, and considering spinorial degrees of freedom might provide an atomic realization of the Einstein-de Haas effect \cite{kawaguchi:080405}.

In view of fermionic dipolar quantum gases, amazing predictions have been made. In the case of homogeneous gases, interesting properties like zero sound \cite{citeulike:6646198,citeulike:5114415}, Berezinskii-Kosterlitz-Thouless phase transition \cite{bruun:245301}, and nematic phases \cite{1367-2630-11-10-103003,fregoso:205301} have been considered, while studies of trapped dipolar gases focus on anisotropic superfluidity in three dimensions \cite{baranov:250403}, fractional quantum Hall states \cite{baranov:070404}, and Wigner crystallization in rotating two-dimensional systems \cite{baranov:200402}.

From the experimental point of view there are different possibilities of realizing dipolar Fermi gases. One of them is to use atoms which have large permanent magnetic dipole moments $m$, such as the isotope $^{53}$Cr of chromium, which has a dipole moment of six Bohr magnetons $\mu_{\rm B}$ and has already been magneto-optically trapped \cite{chicireanu:053406} or the isotope $^{173}$Yb of ytterbium, which has $m = 3~\mu_{\rm B}$ in the $^{3}P_{2}$-state and has already been cooled down to quantum degeneracy \cite{fukuhara:030401}. In addition, recent developments in laser cooling of the isotope $^{66}$Dy of dysprosium, with a magnetic dipole moment of the order $m \sim 10~\mu_{\rm B}$, promises to increase the variety of highly magnetic atomic gases in the quantum degenerate regime \cite{dysp_experiment_two}. A further exciting possibility is displayed by samples of heteronuclear polar molecules. For them, prospects for collisional control through applied electric fields \cite{ticknor:133202} indicate that dipolar gases could be explored all the way from the weak- (collisionless) to the strong-interaction (hydrodynamic) regime, since this may lead to interaction strengths changing by orders of magnitude depending on the applied electric field \cite{1367-2630-11-5-055039}. This is in close analogy to the use of Feshbach resonances to tune the contact interaction to unitarity as has been carried out with success to observe hydrodynamic behavior in the normal phase of atomic Fermi gases \cite{PhysRevLett.91.020402}. Experimentally this is very promising and, recently, $4\times 10^{4}$ fermionic $^{40}$K$^{87}$Rb molecules with an electric dipole moment of about $0.5$ Debye have been brought close to quantum degeneracy by using stimulated Raman adiabatic passage to efficiently convert the molecules into the rovibrational ground state \cite{arXiv:0811.4618}. After that, further progress towards probing quantum degeneracy has been made, by bringing this system into the hyperfine ground state as well \cite{arXiv:0908.3931}. In the meantime, KRb-samples at the temperature $T=1.4~T_{F}$, where $T_{F}$ is the Fermi temperature, have become available, in which thermodynamic measurements led to observation of the anisotropy characteristic of the DDI \cite{citeulike:6565167}.

The first theoretical investigations of such a system were carried out under the assumption of a Gaussian density profile, which is able to capture some of the important features as the aspect ratio, but misses the correct weak-interaction, collisionless limit \cite{PhysRevA.63.033606,PhysRevA.67.025601}. On top of that, other approaches have been pursued including \cite{miyakawa:061603,bruun:245301,1367-2630-11-5-055017} or neglecting \cite{he:031605} the possibility of a deformation of the momentum distribution. However, quantum degenerate heteronuclear polar molecules possess strong dipolar interactions which might lead to a collisional regime combined with an anisotropic momentum distribution. For this reason, we have recently developed a complete theory for normal dipolar Fermi gases in the hydrodynamic regime in the presence of a cylinder-symmetric harmonic trap \cite{ourpaper}. In the present paper we extend our theory to the general case of a triaxial trap, which allows to study important aspects of the physics of dipolar Fermi gases such as the radial quadrupole excitation as well as to sort out the anisotropic effects of the DDI.

In the following we treat one-component fermionic dipolar quantum gases semi-analytically and tacitly assume that the gas is in the hydrodynamic regime. By adapting a variational time-dependent Hartree-Fock method, which was originally developed to study nuclear hydrodynamics \cite{brink,lipparini}, we are able to obtain a complete description of strongly interacting normal dipolar Fermi gases which encompasses their static as well as dynamic properties. The paper is organized as follows. In Section \ref{tdvhf}, we outline the variational formalism of hydrodynamics applied throughout the paper. In Section \ref{eom}, we derive the action governing the dynamics of the system in the case of three different trapping frequencies and extremize it with respect to the widths in spatial and momentum distributions, obtaining, thus, the corresponding equations of motion. Section \ref{dless_v} provides the dimensionless variables which make the physical interpretation of the results more enlightening. Then, in Section \ref{stat_prop}, we derive the equilibrium properties such as the momentum and real space aspect ratios as functions of the dipolar strength and the trap anisotropies. Section \ref{ll_excit} is devoted to the low-lying excitations, where we study the oscillations around the equilibrium. Following that, we address in Section \ref{tof_exp} time-of-flight experiments which represent another fundamental issue in cold atoms physics. In  Section \ref{conc} we present the conclusion where the main results are summarized and further studies of this system are discussed.

\section{\label{tdvhf} Hydrodynamic formulation of Hartree-Fock theory}
Consider a gas containing $N$ harmonically trapped fermionic particles of mass $M$ possessing either electric or magnetic dipole moments, which are polarized in $z$-direction. The Hamilton operator of such a quantum many-particle system is given by
\begin{equation}
H = \sum\limits_{i=1}^{N}\left[-\frac{\hbar^{2}\nabla_{{\bf x}_{i}}^{2}}{2M}  + U_{\rm tr}({\mathbf x}_{i})  \right] + \frac{1}{2}\sum\limits_{i\neq j}^{N}V^{\rm}_{\rm int}({\mathbf x}_{i} -{\mathbf x}_{j}).
\label{HA1}
\end{equation}
The first term represents the kinetic energy, which turns out to be negligible for Bose systems in the Thomas-Fermi regime but is important for Fermi systems, since it provides stability against collapse. In general, the trapping potential is harmonic and reads
\begin{equation}
U_{\rm tr}({\mathbf x}) = \frac{M}{2}\left( \omega^{2}_{x}{x}^{2} + \omega^{2}_{y}{y}^{2} + \omega^{2}_{z}{z}^{2} \right),
\label{HA2}
\end{equation}
where $\omega_{x}$, $\omega_{y}$, and $\omega_{z}$ are different trapping frequencies. The last term in Eq.~(\ref{HA1}) takes care of the interaction and $V^{\rm}_{\rm int}({\mathbf x}_{i} -{\mathbf x}_{j})$ denotes the two-body interaction potential between particles located at ${\mathbf x}_{i}$ and ${\mathbf x}_{j}$.

At very low temperatures, the Pauli exclusion principle prohibits s-wave scattering for identical Fermi particles and long-range interactions, such as the dipole-dipole interaction
\begin{equation}
V_{\rm dd}({\mathbf x}) = \frac{C_{\rm dd}}{4\pi|{\mathbf x}|^{3}}\left[1-3\frac{z^{2}}{|{\mathbf x}|^{2}}\right],
\label{HA3}
\end{equation}
become important. In the case of magnetic dipoles $m$ the DDI is characterized by $C_{\rm dd} = \mu_{0}m^{2}$, with $\mu_{0}$ being the magnetic permeability in vacuum, whereas for electric moments we have $C_{\rm dd} = 4\pi d^{2}$ with the electric dipole moment $d$ expressed in Debyes. For most Fermi gases which have been realized so far, $V_{\rm dd}$ is rather irrelevant and polarization leads to a degenerate non-interacting gas. For polar molecules, however, this is no longer valid: moderate electric fields induce dipole moments which render the DDI (\ref{HA3}) a prominent contribution to the Hamiltonian in Eq.~(\ref{HA1}).

\subsection{Center-of-mass expansion}
In the usual formulation, hydrodynamic studies of degenerate Fermi gases are based on closed equations for the particle density $\rho({\bf x},t)$ and the velocity field ${\bf v}({\bf x_{}},t)$. The dynamic properties of this system are determined by solving the continuity equation and the Euler equation. This set of coupled equations can be obtained from the Boltzmann equation for the phase-space distribution \cite{PhysRevLett.83.5415} or by expanding the equation of motion for the one-body density matrix around the center of mass \cite{amoruso}. Let us explore further the last possibility in order to illustrate some aspects of the method which we apply in the following.

Consider the action governing an $N$-fermion system
\begin{equation}
{\mathcal A} = \int\limits_{t_{1}}^{t_{2}}{\mathrm d}t \langle\Psi|i\hbar\frac{\partial}{\partial t} - H|\Psi\rangle,
\label{action_princ}
\end{equation}
where $|\Psi\rangle$ denotes a Slater determinant built out of one-particle orbitals $\psi_{i}({\bf x}_{},t)$
denoted by
\begin{equation}
\Psi({\bf x}_{1},\cdots,{\bf x}_{N};t) = {\mathrm{ SD}}\left[\psi_{i}({\bf x}_{},t)\right],
\label{sldet}
\end{equation}
with the energy-level index $i$ taking the values $1\le i\le N$.

By extremizing (\ref{action_princ}) with respect to the functions $\psi_{i}({\bf x}_{},t)$ and $\psi^{*}_{i}({\bf x}_{},t)$, one obtains the
Hartree-Fock equations for the one-particle orbitals $\psi_{i}^{*}({\bf x}_{},t)$ and $\psi^{}_{i}({\bf x}_{},t)$. Combining them yields the equation of motion \cite{ring-schuck}
\begin{eqnarray}
i\hbar\frac{\partial \rho({\bf x},\!{\bf x'}\!;t)}{\partial t}\!\! &\!\! =\!\! &\!\! \left[\!\frac{-\hbar^{2}}{2M}\!\!\left(\nabla_{\!{\bf x}}^{2} \!-\!\nabla_{\!{\bf x'}}^{2}\right)\!+\! U_{\rm tr}({\bf x})\!-\!U_{\rm tr}({\bf x'})\right]\!\!\rho({\bf x},\!{\bf x'}\!;t)\nonumber\\
& &+\!\left[\Gamma^{\rm D}({\bf x},t)\!-\!\Gamma^{\rm D}( {\bf x'},t)\right]\rho({\bf x},\!{\bf x'}\!;t)\nonumber\\
&&\!\!\!\!\!\!\!\!\!\!\!\!\!\!\!\!\!\!\!\!\!\!\!\!\!\!\!\!\!\!\!\!\!\!\!\!\!\!+\int\!\!{\mathrm d}^{3}r\left[\Gamma^{\rm E}( {\bf x},{ \bf r};t)\rho({\bf r},{\bf x'};t)\!-\!\Gamma^{\rm E}({\bf r},{\bf x'};t)\rho({\bf x},{ \bf r};t) \right]
\label{hf_eqmot}
\end{eqnarray}
for the one-body density matrix
\begin{eqnarray}
\rho({\bf x},{\bf x'}\!;t) \! \!\!&\!\!\!=\!\!\!&\!\! \! \prod\limits_{i=2}^{N}\!\int\!\! {\mathrm d}^{3} {\bf x}_{i}\! \Psi^{*}\!({\bf x}'\!,{\bf x}_{2},\!\cdots\!,\!{\bf x}_{N};t) \Psi\!({\bf x},{\bf x}_{2},\! \cdots\!,\! {\bf x}_{N};t),\nonumber\\
& = & \sum\limits_{i=1}^{N}\psi_{i}({\bf x}_{},t)\psi^{*}_{i}({\bf x}_{},t).\label{1bdm}
\end{eqnarray}
Here, the direct Hartree term, to which only the diagonal density matrix contributes, reads
\begin{equation}
\Gamma^{\rm D}({\bf x},t) = \int{\mathrm d}^{3}r V_{\rm int}({\bf r},{\bf x})\rho({\bf r},t),
\label{hartree_1}
\end{equation}
while the Fock exchange term, which is given by
\begin{eqnarray}
\Gamma^{\rm E}({\bf x},{\bf x'};t ) & = & - V_{\rm int}({\bf x},{\bf x'})\rho({\bf x},{\bf x'};t),
\label{fock_1}
\end{eqnarray}
also involves off-diagonal elements of the density matrix.

In order to obtain the conservation laws corresponding to the hydrodynamic equations, we perform an expansion around the center-of-mass coordinate ${\bf X}=({\bf x}+{\bf x'})/2$ in powers of the relative coordinate ${\bf s}={\bf x}-{\bf x'}$. In zeroth order in ${\bf s}$, we obtain from (\ref{hf_eqmot}) the continuity equation
\begin{equation}
\frac{\partial \rho({\bf x},t)}{\partial t} + \nabla\cdot{\bf j}({\bf x},t)=0,
\label{conti}
\end{equation}
with the particle density $\rho({\bf x},t)=\rho({\bf x},{\bf x};t)$ and the current density
\begin{equation}
{\bf j}({ {\bf x},t})=\frac{\hbar}{2Mi}\left(\nabla_{{\bf x}}-\nabla_{{\bf x'}} \right)\rho({\bf x},{\bf x'};t)\bigg|_{{\bf x'}={\bf x}}.
\label{curr_dens}
\end{equation}
The first order in ${\bf s}$ yields from (\ref{hf_eqmot}) the Euler equation
\begin{eqnarray}
M\frac{\partial j_{i}({\bf x},t)}{\partial t} & = & -\nabla_{{\bf x}_{j}}\Pi_{ij}^{0}({\bf x},t)-\rho({\bf x},t)\nabla_{{\bf x}_{i}}U({\bf x})\nonumber\\
& &\!\!\!\!\!\!\!\!\!\!\!\!\!\!\!\!\!\!\!\!\!\!\!\!\!\!\!\!\!\!\!\!\!\!\!\!\!\!\!\!-{\rho({\bf x},\!t)}\nabla_{{\bf x}_{i}}\Gamma^{\rm D}\!({\bf x},\!t)\!+\!\!\!\int\!\! {\mathrm d}^{3}x' {\rho_{}({\bf x},\!{\bf x'}\!;\!t)\rho_{}({\bf x'}\!,\!{\bf x};\!t)}\nabla_{\bf x}V_{\rm int}({\bf x},{\bf x'}),\nonumber\\
\label{euler}
\end{eqnarray}
with the non-interacting kinetic stress tensor
\begin{equation}
\Pi_{ij}^{0}({\bf x},t) = -\frac{\hbar^{2}}{M}\frac{\left(\nabla_{{\bf x}}\!-\!\nabla_{{\bf x'}} \right)_{i}}{2}\frac{\left(\nabla_{{\bf x}}\!-\!\nabla_{{\bf x'}} \right)_{j}}{2}\rho({\bf x},{\bf x'};t)\bigg|_{{\bf x'}={\bf x}}.
\end{equation}
Introducing the velocity field ${\bf v}({\bf x},t)={\bf j}({ {\bf x},t})/\rho({\bf x},t)$ and assuming that the trapping potential $U({\bf x})$ is sufficiently smooth, the kinetic stress tensor takes the form \cite{amoruso}
\begin{equation}
\Pi_{ij}^{0}({\bf x},t) = \delta_{ij}P^{0}({\bf x},t) + M\rho({\bf x},t){v}_{i}({\bf x},t){v}_{j}({\bf x},t),
\end{equation}
where the pressure $P^{0}({\bf x},t)$ obeys some equation of state $P^{0}({\bf x},t)=F\left(\rho({\bf x},t)\right)$.

In case of an irrotational flow, where the circulation of the velocity field vanishes due to $\nabla\times{\bf v}={\bf 0}$, the Euler equation (\ref{euler}) can be rewritten in the form
\begin{eqnarray}
M\frac{\rm d {\bf v}({\bf x},t)}{\rm d t} & = & -\nabla\left[\int\limits^{\rho({\bf x},t)} {\mathrm d}\rho'\frac{F'(\rho')}{\rho'}+U({\bf x})+\Gamma^{\rm D}({\bf x},t)\right]\nonumber\\
& & \!\!\!\!\!\!\!\!\!\!\!\!\!\!\!\!+\int\!\! {\mathrm d}^{3}x' \frac{\rho_{}({\bf x},\!{\bf x'}\!;\!t)\rho_{}({\bf x'}\!,\!{\bf x};\!t)}{\rho({\bf x},t)}\nabla_{\bf x}V_{\rm int}({\bf x},{\bf x'})
\label{eulers2}
\end{eqnarray}
with the transport derivative ${\rm d }/{\rm d t}={\partial}/{\partial t} + {\bf v}\cdot \nabla$. Now the effect of the exchange term of the non-local interaction potential (\ref{HA3}) becomes clear: it breaks the conservation of the circulation of the velocity field ${\bf v}({\bf x},t)$ and Kelvin's theorem does not hold although we consider an irrotational flow. This obvious contradiction is a consequence of the fact that it is {\sl a priori} not possible to describe the exchange correlations in terms of density fluctuations alone, i.e., fluctuations of the diagonal part of the one-particle density matrix. Of course, the true exchange correlation is a function of the density alone, as a consequence of the Kohn theorem \cite{PhysRev.136.B864}, and circulation is conserved. Thus, due the presence of the Fock exchange term, the hydrodynamic treatment commonly used for dipolar BECs \cite{PhysRevLett.92.250401} cannot immediately be applied to degenerate dipolar Fermi gases. For this reason, we propose another approach which preserves the influence of the non-diagonal part of the one-particle density matrix \cite{ourpaper}, yet assures the conservation of the  velocity circulation.

\subsection{Common-phase approach}

In this section we discuss the variational time-de\-pend\-ent approach for a general two-particle interaction potential, which will lead to a unified formalism for elucidating the hydrodynamic properties of normal dipolar Fermi gases.

In order to study the collective motion of the gas, we employ a crucial approximation for the one-particle orbitals $\psi_{i}({\bf x},t)$, namely that they all have the same phase
\begin{equation}
\psi_{i}({\bf x},t) = e^{iM\chi({\bf x},t)/\hbar} |\psi_{i}({\bf x},t)|.
\label{psi_ansa}
\end{equation}
This approximation was introduced before in the context of nuclear hydrodynamics \cite{brink} and is commonly used in hydrodynamic studies (see, for instance, Ref.~\cite{lipparini}). The orbitals $|\psi_{i}({\bf x},t)|$ are invariant under time reversion and are, therefore, called time-even.

From Eq.~(\ref{psi_ansa}) and the definition of a time-even Slater determinant $\Psi_{0}({\bf x}_{1},\cdots,{\bf x}_{N};t)= {\mathrm{ SD}}\left[|\psi_{i}({\bf x}_{},t)|\right]$ we obtain
\begin{equation}
\Psi\!({\bf x}_{1},\!\cdots\!,{\bf x}_{N};t)\! =\! e^{i\!\frac{M}{\hbar}\left[\chi({\bf x}_{1}\!,t)+\cdots+\chi({\bf x}_{N}\!,t)\right]}\Psi_{0}\!({\bf x}_{1},\!\cdots\!,{\bf x}_{N};t).
\label{}
\end{equation}
Thus, the one-body density matrix (\ref{1bdm})
reduces to
\begin{equation}
\rho({\bf x},{\bf x'};t) =  e^{i\frac{M}{\hbar}\left[\chi({\bf x},t)-\chi({\bf x'},t)\right]}\rho_{0}({\bf x},{\bf x'};t),
\label{rho_0}
\end{equation}
with $\rho_{0}({\bf x},{\bf x'};t)$ being a time-even one-body density matrix. At this point it becomes more evident that the present method resembles that of the collective coordinates applied for fermions, as mentioned in chapter 16 of Ref. \cite{pethick}.

Now the current density, defined in Eq.~(\ref{curr_dens}), becomes ${\bf j}({ {\bf x},t})=\rho_{0}({\bf x},t)\nabla \chi({\bf x},t)$, allowing for the identification of $\chi({\bf x},t)$ as the potential of the velocity field ${\mathbf v}({\bf x},t)$.

With these definitions the action (\ref{action_princ}) reduces to
\begin{eqnarray}
{\cal A} & \!\!=\!\! &  \!-\!M\!\!\int\limits_{t_{1}}^{t_{2}}\!\!{\mathrm d}t  \!\!\!\int\!\! {\mathrm d}^{3}x\left\{ \dot{\chi}({\bf x},t)\rho_{0}({\bf x},t)
+\frac{\rho_{0}({\bf x},t)}{2}\left[\nabla\chi({\bf x},t)\right] ^{2} \right\}\nonumber\\
& & -  \int\limits_{t_{1}}^{t_{2}}{\mathrm d}t \langle\Psi_{0}|{H}|\Psi_{0}\rangle.
\label{action_rho0}
\end{eqnarray}
The first two terms concern the dynamical properties of the system and will be shown to give rise to the time derivatives in the equations of motion. Notice that integrating the first term by parts shows that the common phase $\chi({\bf x},t)$ can be seen as the momentum conjugate to coordinate $\rho_{0}({\bf x},t)$, which represents the particle density. The second term describes the energy associated with the movement, i.e., the flow energy \cite{pethick}, given by
\begin{equation}
E_{\rm flow}(t) = \frac{M}{2}\int {\mathrm d}^{3} x \rho_{0}({\bf x},t)\left[\nabla\chi({\bf x},t)\right]^{2}.
\label{flow_ener}
\end{equation}

The last term of Eq.~(\ref{action_rho0}), i.e., $\langle\Psi_{0}|H|\Psi_{0}\rangle$, consists in total of three contributions
\begin{equation}
\langle\Psi_{0}|H|\Psi_{0}\rangle = \langle\Psi_{0}|H_{\rm kin}|\Psi_{0}\rangle + \langle\Psi_{0}|H_{\rm tr} |\Psi_{0}\rangle+ \langle\Psi_{0}|H_{\rm int}|\Psi_{0}\rangle.
\label{hamil}
\end{equation}
The first one is the expectation value of the kinetic energy operator with respect to $|\Psi_{0}\rangle$ and gives rise to the Fermi pressure:
\begin{equation}
E_{\rm kin} (t)=  \frac{-\hbar^{2}}{2M}\int {\mathrm d}^{3}x{\left(\nabla_{\!\!{\bf x}}\!\!-\!\!\nabla_{\!\!{\bf x'}} \right)\cdot\left(\nabla_{\!\!{\bf x}}\!\!-\!\!\nabla_{\!\!{\bf x'}} \right)}\rho_{0}({\bf x},{\bf x'};t)\bigg|_{{\bf x'}={\bf x}}.\label{kin_0}
\end{equation}
Notice that the total kinetic energy is given by $E_{\rm flow} + \langle\Psi_{0}|H_{\rm kin}|\Psi_{0}\rangle$. For simplicity, the kinetic energy in the static case, i.e., $\langle\Psi_{0}|H_{\rm kin}|\Psi_{0}\rangle$, will be referred to as Fermi pressure or simply kinetic energy. The second term in Eq.~(\ref{hamil}) represents the energy of the external trapping potential
\begin{equation}
E_{\rm tr} (t)= \int {{\mathrm d}^{3}x} \,\rho_{0}({\bf x},t)U_{\rm tr}({\mathbf x}).\label{trap_0}
\end{equation}

The interaction energy, given by the third term in Eq.~(\ref{hamil}), contains both the direct and the exchange mean-field terms $\langle\Psi_{0}|H_{\rm int}|\Psi_{0}\rangle =E_{\rm}^{\rm D} + E_{\rm}^{\rm E}$. The direct contribution is given by
\begin{eqnarray}
\!\!E_{\rm}^{\rm D}\!(t) & \!\!\!= \!\!\!&  \frac{1}{2}\!\!\int\!\! {\mathrm d}^{3}x{\mathrm d}^{3}x' V_{\rm int}({\bf x},{\bf x'};t)\rho_{0}({\bf x},{\bf x};t)\rho_{0}({\bf x'},{\bf x'};t),\nonumber\\ \label{direc_0}
\end{eqnarray}
while the exchange part reads
\begin{eqnarray}
\!\!E_{\rm}^{\rm E}\! (t)&\!\!\! = \!\!\!&-\frac{1}{2}\!\!\int\!\!\! {\mathrm d}^{3}x{\mathrm d}^{3}x' V_{\rm int}({\bf x},{\bf x'};t)\rho_{0}({\bf x},{\bf x'};t)\rho_{0}({\bf x'},{\bf x};t).\nonumber\\ \label{exchang_0}
\label{hamil_int}
\end{eqnarray}
Of course, if the interaction energy $\langle\Psi_{0}|H_{\rm int}|\Psi_{0}\rangle$ would be a functional of the particle density $\rho_{0}({\bf x},t)$ alone, conservation laws corresponding to the continuity equation and the Euler equation could be immediately derived by functionally extremizing the action (\ref{action_rho0}) with respect to the phase $\chi({\bf x},t)$ and the density $\rho_{0}({\bf x},t)$, respectively. In the present case, however, one has to extremize with respect to the full time-even one-body density matrix $\rho_{0}({\bf x},{\bf x'};t)$. It turns out that the continuity equation remains unchanged
\begin{eqnarray}
\frac{\partial \rho_{0}({\bf x},t)}{\partial t} & = & -\nabla\cdot\left[\rho_{0}({\bf x},t){\bf v}({\bf x},t)\right].
\label{conti_0}
\end{eqnarray}
The corresponding Euler equation reads, formally,
\begin{equation}
M\frac{\rm d {\bf v}({\bf x},t)}{\rm d t} = -\nabla\left[\int {\mathrm d}^{3}x'\frac{\delta \langle\Psi_{0}|H_{\rm }|\Psi_{0}\rangle}{\delta \rho_{0}({\bf x},{\bf x'};t)}\right],
\label{euler_2nd}
\end{equation}
so that the proposed approach is circulation conserving.

\subsection{Wigner phase space}

In the preceeding section we have derived a set of equations which could be applied to study the hydrodynamic excitations of a dipolar Fermi gas. Nevertheless, the equations (\ref{conti_0}) and (\ref{euler_2nd}) are not yet closed due to the lack of knowledge of the non-diagonal terms of the density matrix. Therefore, we resort to a simpler procedure: we extremize action (\ref{action_rho0}) with respect to appropriate variational quantities, which bear information on both the diagonal as well as the non-diagonal part of the interaction. Of course, we do loose information in this process because our variational approach may not be as precise as the solution of the complicated equations (\ref{hf_eqmot}), but, on the other hand, it gives access to both the statical and dynamical properties of dipolar Fermi gases beyond the perturbative regime in a quite simple and clear way.

In order to calculate each of the terms in the action (\ref{action_rho0}), we change to the Wigner representation of the time-even one-body density matrix, which is defined according to
\begin{eqnarray}
\nu_{0}\left({\mathbf x},{\mathbf k};t\right) & = & \int {\mathrm d}^{3}s\,\rho_{0}\left({\mathbf x}+\frac{\mathbf s}{2},{\mathbf x}-\frac{\mathbf s}{2};t\right)\,e^{-i{\mathbf k}\cdot{\mathbf s}}.
\label{wig_def}
\end{eqnarray}
The inverse transformation reads
\begin{eqnarray}
\!\rho_{0}({\mathbf x}_{},{\mathbf x'}_{};t) \!\!& =\!\! & \int \frac{{\mathrm d}^{3}k}{(2\pi)^{3}}\,\nu_{0}\!\left(\frac{{\mathbf x}+{\mathbf x'}}{2},{\mathbf k};t\right)\!e^{i{\mathbf k}\cdot({\mathbf x}-{\mathbf x'})}.
\end{eqnarray}
In the Wigner representation all quantities of interest can be expressed in terms of the Wigner function (\ref{wig_def}). For example, the particle density is given by
\begin{equation}
\rho_{0}({\mathbf x}_{},t) = \rho_{0}({\mathbf x}_{},{\mathbf x}_{};t) = \int \frac{{\mathrm d}^{3}k}{(2\pi)^{3}}\,\nu_{0}\left({{\mathbf x}}{},{\mathbf k};t\right),
\label{part_dens}
\end{equation}
and the momentum distribution is obtained via
\begin{equation}
\rho_{0}({\mathbf k}_{},t) = \int \frac{{\mathrm d}^{3}x}{(2\pi)^{3}}\,\nu_{0}\left({{\mathbf x}}{},{\mathbf k};t\right).
\label{mom_dist}
\end{equation}
With the help of these quantities, the kinetic energy (\ref{kin_0}) and the trapping (\ref{trap_0}) energy can be written as
\begin{eqnarray}
E_{\rm kin} (t)& = & \int \frac{{\mathrm d}^{3}x{\mathrm d}^{3}k} {(2\pi)^{3}}\,\nu_{0}\left({{\mathbf x}}{},{\mathbf k};t\right)\frac{\hbar^{2}{\mathbf k}^{2}}{2M}\label{E_kin},\\
E_{\rm tr} (t)& = & \int \frac{{\mathrm d}^{3}x{\mathrm d}^{3}k} {(2\pi)^{3}}\,\nu_{0}\left({{\mathbf x}}{},{\mathbf k};t\right)U_{\rm tr}({\mathbf x}),\label{E_trap}
\end{eqnarray}
respectively. Accordingly, the direct term in (\ref{direc_0}), which represents the mean-field dipolar potential energy, reads
\begin{eqnarray}
E_{\rm dd}^{\rm D}\!(t) \!\!\! & = &\!\!\!\!\! \int\!\! \frac{{\mathrm d}^{3}x{\mathrm d}^{3}k{\mathrm d}^{3}x'{\mathrm d}^{3}k'}{2(2\pi)^{6}} 
\nu_{0}\!\left({{\mathbf x}},{\mathbf k};\!t\right)\!V_{\!\rm dd}({\mathbf x}\!-\!{\mathbf x'})\nu_{0}\!\left({{\mathbf x'}}\!,{\mathbf k'}\!;\!t\right)\!.\nonumber\\
\label{int_dir}
\end{eqnarray}
This term is determined by the particle density (\ref{part_dens}) alone and was first considered to analyze the equilibrium \cite{PhysRevA.63.033606} and the dynamical \cite{PhysRevA.67.025601} properties of a cylinder-symmetric system by adopting an isotropic momentum distribution and a Gaussian trial particle density.

In contrast, the exchange interaction term (\ref{exchang_0}), given by
\begin{eqnarray}
E_{\rm dd}^{\rm E}(t) \!\! & = &\!\! -\!\!\int\!\! \frac{{\mathrm d}^{3}\!X{\mathrm d}^{3}k{\mathrm d}^{3}\!s{\mathrm d}^{3}k'}{2(2\pi)^{6}} 
\nu_{0}\!\left({{\mathbf X}},{\mathbf k};t\right)\!V_{\rm dd}({\mathbf s})\nu_{0}\!\left({{\mathbf X}},{\mathbf k'};t\right)\nonumber\\
\!\!\! & &\!\!\!\times e^{i{\mathbf s}\cdot({\mathbf k}-{\mathbf k'})}, \, 
\label{int_exc}
\end{eqnarray}
is rather linked to the momentum distribution (\ref{mom_dist}) and vanishes if it is isotropic. The importance of this term in dipolar Fermi gases was only recently recognized in Ref.~\cite{miyakawa:061603}, where it was shown to lead to a deformed momentum distribution. Following this important investigation, further effects have been studied by taking this term into account, like its influence on the dynamical properties of cylindrically trapped systems in the collisionless regime \cite{1367-2630-11-5-055017}. For homogeneous gases, zero-sound dynamics \cite{citeulike:6646198,citeulike:5114415} as well as quantum phase transitions in two dimensions \cite{bruun:245301} have been investigated by considering also this exchange contribution (\ref{int_exc}).

In the following we extend the static, semi-classical theory of dipolar Fermi gases to a dynamical theory in the hydrodynamic regime by including the Fock exchange term in a natural way. Actually, as is clear from its derivation, this theory can also be successfully applied to fermionic systems with other types of long-range interactions.

\section{\label{eom}Equations of motion}

In order to study the trapped dipolar Fermi gas, we adopt for the common phase the harmonic ansatz
\begin{eqnarray}
\chi({\bf x},t) & = & \frac{1}{2}\left[\alpha_{x}(t)x_{}^{2} + \alpha_{y}(t)y_{}^{2} +\alpha_{z}(t)z_{}^{2}\right],
\label{ansatz_phase}
\end{eqnarray}
which is able to capture different excitation modes by specifying the form of the potential for the particle flow. Furthermore, we use an ansatz for the Wigner function which resembles that of a non-interacting Fermi gas in the semiclassical approximation. With this we cope with the main effect of the DDI that the gas is stretched in the direction of the polarization. This ansatz is a generalization of the one presented in Ref.~\cite{miyakawa:061603}, which has the form of the low-temperature limit of the Fermi-Dirac distribution
\begin{eqnarray}
\!\!\!\!\!\!\!\nu_{0}\left({{\mathbf x}},{\mathbf k};t\right) & = & \Theta\left(1-\sum\limits_{i}\frac{x^{2}_{i}}{R_{i}(t)^{2}}-\sum\limits_{i}\frac{k_{i}^{2}}{K_{i}(t)^{2}} \right),
\label{ansatz_wfunc}
\end{eqnarray}
with $\Theta(x)$ denoting the Heaviside step function.
According to Eq.~(\ref{ansatz_wfunc}), the parameters $R_{i}$ and $K_{i}$ represent the largest extension in the $i$-th direction of the density and momentum distribution, respectively. They will, therefore, be called the Thomas-Fermi (TF) radius and Fermi momentum in the $i$-th direction, respectively.

Now we are in a position to evaluate the action (\ref{action_rho0}) as a function of the variational parameters. Introducing additionally the chemical potential $\mu$ as the Lagrange parameter, which is responsible for particle number conservation, the action reads
\begin{eqnarray}
{\mathcal A} & = & -\int\limits_{t_{1}}^{t_{2}}{\mathrm d}t\frac{\overline{R}^{3}\overline{K}^{3}}{3\cdot2^{7}}\left\{\frac{M}{2}\sum\limits_{i}\left[\dot{\alpha}_{i}R_{i}^{2}+\alpha_{i}^{2}R_{i}^{2}+\omega_{i}^{2}R_{i}^{2}\right]\right.\nonumber\\
\!\!\!\!\!\!&  &\!\!\!\!\!\!\!\!\!\!\!\!+\left.\sum\limits_{i}\frac{\hbar^{2}K_{i}^{2}}{2M}-c_{0}\overline{K}^{3} \left[f\left(\frac{R_{x}}{R_{z}},\frac{R_{y}}{R_{z}}\right)-f\left(\frac{K_{z}}{K_{x}},\frac{K_{z}}{K_{y}}\right)\right]\right\}\nonumber\\
& & -\int\limits_{t_{1}}^{t_{2}}{\mathrm d}t\, \mu(t)\left(\frac{\overline{R}^{3}\overline{K}^{3}}{48}-N\right),
\label{final_action}
\end{eqnarray}
where $\overline{\bullet}$ denotes geometrical average and the constant $c_{0}$ is given by
\begin{equation}
c_{0} = \frac{2^{10}C_{\rm dd}}{3^{4}\cdot5\cdot7\cdot\pi^{3}}\approx 0.0116~C_{\rm dd}.
\end{equation}
\begin{figure}
\includegraphics[scale=.5]{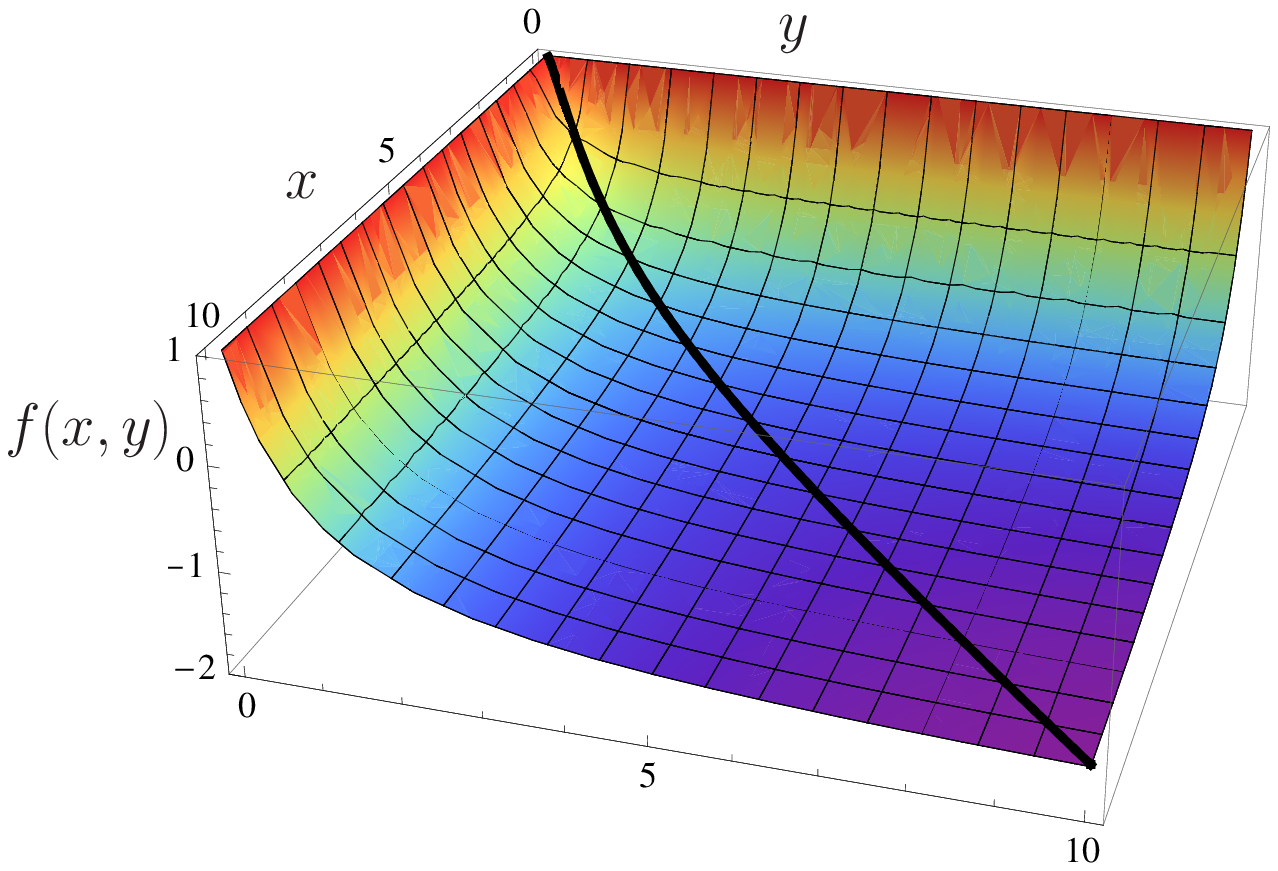}
\quad\includegraphics[scale=.8]{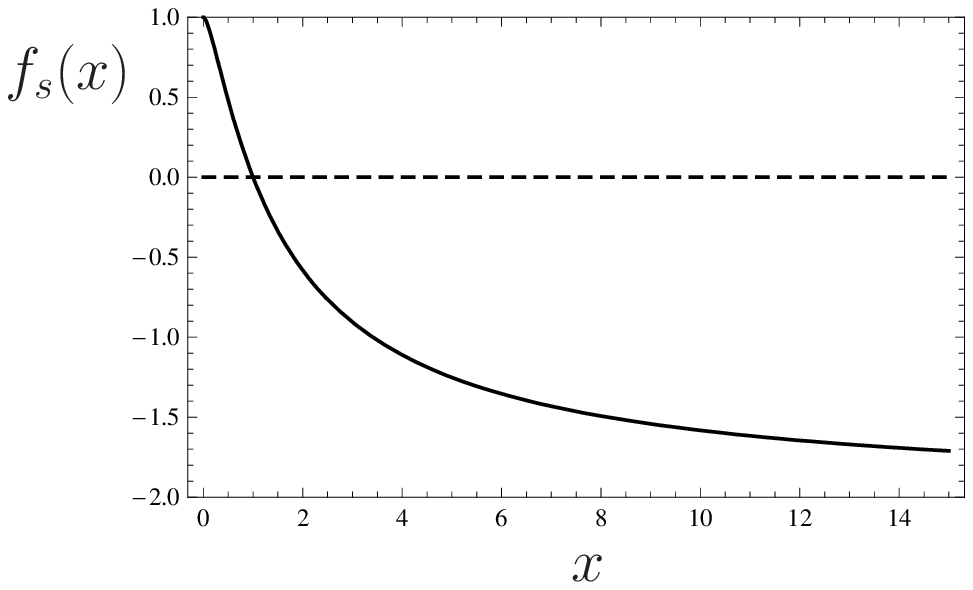}
\caption{(Color Online) Top: Anisotropy function $f(x,y)$ from (\ref{aniso_func}), which is bounded between $1$ for small values of either $x$ or $y$ and $-2$ for large values of both arguments. Notice the symmetry $f(x,y)=f(y,x)$ and that $f(x,y)$ reduces to $f_{s}(x)$ (black curve) in the case of cylindrical symmetry \cite{glaum:023604,giovanazzi:013621}. Bottom: $f_{s}(x)$ as a function of $x$. Notice that $f_{s}(x)$ changes its sign at $x=1$.}
\label{fig_1}
\end{figure}
The DDI is reflected in the anisotropy function $f(x,y)$, which is defined as (see Fig.~\ref{fig_1})
\begin{equation}
f(x,y) = 1 + 3xy\frac{E(\varphi,q)-F(\varphi,q)}{(1-y^{2})\sqrt{1-x^{2}}},
\label{aniso_func}
\end{equation}
where $F(\varphi,q)$ and $E(\varphi,q)$ are the elliptic integrals of the first and second kind, respectively, with $\varphi=\arcsin\sqrt{1-x^{2}}$ and $q^{2}=(1-y^{2})/(1-x^{2})$. This function has often appeared in the literature of dipolar Bose-Einstein condensates \cite{glaum:023604,giovanazzi:013621}. Notice that $f(x,y)$ is bounded between $1$ and $-2$ passing through $0$ at $x=y=1$. This reflects the fact that the DDI is both partially attractive and partially repulsive, depending on whether the dipoles are head-to-tail or side-by-side to one another, respectively. For polarization along the symmetry axis, therefore, the interaction is dominantly attractive in cigar- (prolate) and repulsive in pancake-shaped (oblate) systems. We remark that this is not the only possible way to define the anisotropic dipolar function \cite{miyakawa:061603}. The motivation for our choice is twofold. From the mathematical point of view, much is known about the function (\ref{aniso_func}) due to the extensive research on dipolar BEC's. For example, a good amount of technical information can be found in the appendix of Ref.~\cite{giovanazzi:013621}. In addition, it is physically appealing to state that the anisotropy of harmonically trapped, dipolar particles is determined by one and the same function no matter if they are of bosonic or fermionic nature.

In the following, we will use the indexes $1$ and $2$ to denote a derivative with respect to the first and second argument. Note that, whereas $f(x,y)$ is symmetric with respect to exchanging the first and second variables, this is not the case for the functions $f_{1}(x,y)$ and $f_{2}(x,y)$. Furthermore, in the case of $x=y$, the anisotropy function $f(x,y)$ reduces to \cite{PhysRevA.63.053607,glaum:080407,PhysRevLett.92.250401}
\begin{equation}
f_{s}(x) =  \frac{1 + 2x^{2} - 3 x^{2}\,\Xi(x)}{1-x^{2}},
\end{equation}
together with the abbreviation
\begin{equation}
\Xi(x)  \equiv  \begin{cases}
\frac{1}{\sqrt{1-x^{2}}}\tanh^{-1}\sqrt{1-x^{2}}; & 0\leq x<1\\
\frac{1}{\sqrt{x^{2}-1}}\tan^{-1}\sqrt{x^{2}-1}; & x\ge1
\end{cases}.
\end{equation}

Inspecting the action (\ref{final_action}), one perceives that the anisotropy function occurs twice. The first time as a function of the TF radii $R_{i}$, due to the direct term, and the second time as a function of the Fermi momenta $K_{i}$, due to the exchange term. Since $f(1,1)$ vanishes, this term only contributes in the case of deformed Fermi surfaces. In other words, the absence of this term would lead immediately to a spherical momentum distribution.

The equations of motion follow from extremizing the action (\ref{final_action}) with respect to all variational parameters $\alpha_{i}, R_{i}, K_{i}$ as well as the Lagrange parameter $\mu$. The latter assures particle number conservation
\begin{equation}
\overline{R}^{3}\overline{K}^{3}=48N
\label{part_num_conse}
\end{equation}
and is explicitly given by
\begin{equation}
{\mu} =\frac{1}{3} \sum\limits_{i}^{}\frac{\hbar^{2}{K}_{i}^{2}}{2M}-\frac{21c_{0}N}{\overline{{R}}^{3}}\!\left[f_{}\!\left(\frac{{R}_{x}}{{R}_{z}},\frac{{R}_{y}}{{R}_{z}}\right)-\!f_{}\!\left(\frac{{K}_{z}}{{K}_{x}},\frac{{K}_{z}}{{K}_{y}} \right)\right].
\label{mu_unitfull}
\end{equation}

After some simple though tedious algebra, one obtains the following equations for the Fermi momenta
\begin{eqnarray}
\frac{\hbar^{2}K_{x}^{2}}{2M} & = & \frac{1}{3}\sum\limits_{i}^{}\frac{\hbar^{2}K_{i}^{2}}{2M} + \frac{48Nc_{0}} {2R_{x}^{}R_{y}^{}R_{z}^{}}\frac{K_{z}}{K_{x}}f_{1}\left(\frac{K_{z}}{K_{x}},\frac{K_{z}}{K_{y}}\right),\nonumber\\
\frac{\hbar^{2}K_{y}^{2}}{2M} & = & \frac{1}{3}\sum\limits_{i}^{}\frac{\hbar^{2}K_{i}^{2}}{2M} + \frac{48Nc_{0}} {2R_{x}^{}R_{y}^{}R_{z}^{}}\frac{K_{z}}{K_{y}}f_{2}\left(\frac{K_{z}}{K_{x}},\frac{K_{z}}{K_{y}}\right),\nonumber\\
\frac{\hbar^{2}K_{z}^{2}}{2M} & = & \frac{1}{3}\sum\limits_{i}^{}\frac{\hbar^{2}K_{i}^{2}}{2M} - \frac{48Nc_{0}} {2R_{x}^{}R_{y}^{}R_{z}^{}}\frac{K_{z}}{K_{x}}f_{1}\left(\frac{K_{z}}{K_{x}},\frac{K_{z}}{K_{y}}\right)\nonumber\\
&&- \frac{48Nc_{0}} {2R_{x}^{}R_{y}^{}R_{z}^{}}\frac{K_{z}}{K_{y}}f_{2}\left(\frac{K_{z}}{K_{x}},\frac{K_{z}}{K_{y}}\right).
\label{k_dynamics}
\end{eqnarray}
These equations are clearly redundant, so we drop the third of them. Together with the condition for particle conservation (\ref{part_num_conse}), they determine the Fermi momenta $K_{i}$ as functions of the TF radii $R_{i}$, so that we have three independent equations to solve for three variables.

The equations of motion for the variational parameters $\alpha_{i}$ are simply given by
\begin{equation}
{\alpha_{i}} = \frac{\dot{R_{i}}}{R_{i}},
\label{alpha_rel}
\end{equation}
and are used to derive the equations of motion for the TF radii:
\begin{eqnarray}
\ddot{R_{i}} & = & -\omega_{i}^{2}R_{i} + \sum\limits_{j}^{}\frac{\hbar^{2}K_{j}^{2}}{3M^{2}R_{i}} -\frac{48Nc_{0}}{Mc_{\rm d}}Q_{i}\left(\bf R, K\right).\nonumber\\
\label{tf_eqs}
\end{eqnarray}
Here the auxiliary functions are given by
\begin{eqnarray}
Q_{x}\left(\bf r, k \right)\!\!& = \!\! & \frac{c_{\rm d}}{{x}^{2}{y}^{}{z}^{}} \!\!\left[f_{}\!\left(\frac{{x}}{{z}},\frac{{y}}{{z}}\right)\!-\!\frac{{x}}{{z}}f_{1}\!\left(\frac{{x}}{{z}},\frac{{y}}{{z}}\right)\!-\!f_{}\!\left(\frac{{k}_{z}}{{k}_{x}},\frac{{k}_{z}}{{k}_{y}}\right)\right],\label{subeqn1:Q_functions}\nonumber\\
Q_{y}\left(\bf r, k  \right)\!\!& = \!\! & \frac{c_{\rm d}}{{x}^{}{y}^{2}{z}^{}} \!\!\left[f_{}\!\left(\frac{{x}}{{z}},\frac{{y}}{{z}}\right)\!-\!\frac{{y}}{{z}}f_{2}\!\left(\frac{{x}}{{z}},\frac{{y}}{{z}}\right)\!-\!f_{}\!\left(\frac{{k}_{z}}{{k}_{x}},\frac{{k}_{z}}{{k}_{x}}\right)\right],\label{subeqn2:Q_functions}\nonumber\\
Q_{z}\left(\bf r, k \right)\!\!& = \!\! & \frac{c_{\rm d}}{{x}^{}{y}^{}{z}^{2}} \!\!\left[f_{}\!\left(\frac{{x}}{{z}},\frac{{y}}{{z}}\right)\!+\!\frac{{x}}{{z}}f_{1}\!\left(\frac{{x}}{{z}},\frac{{y}}{{z}}\right)\!+\!\frac{{y}}{{z}}f_{2}\!\left(\frac{{x}}{{z}},\frac{{y}}{{z}}\right)\right.\nonumber\\& & \left.
\!-\!f_{}\!\left(\frac{{k}_{z}}{{k}_{x}},\frac{{k}_{z}}{{k}_{x}} \right)\right],
\label{subeqn3:Q_functions}
\end{eqnarray}
where the numerical constant $c_{\rm d}$ reads
\begin{equation}
c_{\rm d} = \frac{2^{\frac{38}{3}}}{3^{\frac{23}{6}}\cdot5\cdot7\cdot\pi^{2}}\approx0.2791.
\end{equation}

The first term on the right-hand side of equations (\ref{tf_eqs}) accounts for the harmonic trap, the second is due to the Fermi pressure, and the third represents the DDI contribution, which will be discussed in more detail in the next section.

Having collected the equations of motion for all the variables, we can attempt to interpret Eqs.~(\ref{part_num_conse}), (\ref{k_dynamics}), and (\ref{tf_eqs}) physically. In the case of a spherically symmetric momentum distribution one could neglect the exchange term and set to zero all terms which involve $f_{}({K_{z}}/{K_{x}},{K_{z}}/{K_{y}})$ and its derivatives. Thus, we could solve (\ref{part_num_conse}) and (\ref{k_dynamics}) for the Fermi momenta and obtain
\begin{equation}
K_{x}=K_{y}=K_{z}=K_{F} = \frac{\sqrt[3]{48N}}{\overline{R}_{}}.
\label{nonin_fm}
\end{equation}
Inserting this result in Eqs.~(\ref{tf_eqs}), we, then, would have a set of equations of motion for the TF radii which stem from a potential $V_{\rm }(R_{x},R_{y},R_{z})$. The problem would be reduced to study the movement of a fictitious particle under the influence of this potential. Due to the presence of the Fock term, however, it is not possible to solve Eqs.~(\ref{part_num_conse}) and (\ref{k_dynamics}) directly, so they have to be solved simultaneously with Eqs.~(\ref{tf_eqs}). Thus, we conclude that the exchange term modifies the constraints in an anisotropic manner such that one has to give up the notion of an underlying potential $V_{\rm }(R_{x},R_{y},R_{z})$. Furthermore, it is the presence of the Fock exchange term, as first pointed in \cite{miyakawa:061603}, which deforms the Fermi sphere into an ellipsoid in the case of a cylinder-symmetric trap. As we shall see below, this deformation turns out to remain ellipsoidal for triaxial traps and plays an important role for determining both equilibrium and dynamical properties of the system.

\section{\label{dless_v} Dimensionless variables: Cylindrical symmetry of the momentum distribution}
Before we explore the physical consequences of the equations of motion for a trapped dipolar Fermi gas, let us briefly discuss the non-interacting case, which will provide us with adequate units for the quantities of interest throughout this work.

Denoting the Fermi energy of a non-interacting trapped Fermi gas by $E_{F}$, its chemical potential takes the form
\begin{equation}
\mu^{(0)} = E_{F} = \hbar \overline{\omega}\left(6N \right)^{\frac{1}{3}},
\label{non_mu}
\end{equation}
and the Fermi radii and momentum read, respectively,
\begin{equation}
R_{i}^{(0)} = \sqrt{\frac{2E_{F}}{M\omega_{i}^{2}}};\qquad K_{F} = \sqrt{\frac{2ME_{F}}{\hbar^{2}}}.
\label{non_tf_r_f}
\end{equation}

This motivates to express the TF radii $R_{i}$ in units of $R_{i}^{(0)}$ and the Fermi momenta $K_{i}$ in units of $K_{F}$. Defining $\tilde{K}_{i}\equiv{K_{i}}/{K_{F}}$ and $\tilde{R}_{i}\equiv{R_{i}}/{R_{i}^{(0)}}$, the condition for the particle number conservation (\ref{part_num_conse}) reduces to
\begin{equation}
\overline{\tilde{R}_{}}^{3}\,\overline{\tilde{K}_{}}^{3} = 1.
\label{dim_less_part_num_conse}
\end{equation}
Thus, the equations of motion for the TF radii in the dimensionless notation will be written in terms of the ratios
\begin{equation}
\frac{R_{i}}{R_{j}} = \frac{\tilde{R}_{i}}{\tilde{R}_{j}}\frac{\omega_{j}}{\omega_{i}},
\label{Rs_dim_less}
\end{equation}
outlining the role played by the trap frequency ratios $\lambda_{x} = {\omega_{z}}/{\omega_{x}}$ and $\lambda_{y} = {\omega_{z}}/{\omega_{y}}$.

Before solving the equations of motion, we can already obtain important information by considering the symmetries of the total energy in the static case, i.e., with the velocity potential $\chi$ set to zero. As a function of the variational parameters, the energy then reads
\begin{eqnarray}
\!\!\!\!\!\!\!\!\!\!\frac{E}{N E_{F}} &\!\! = \!\!& \frac{1} {8}\left\{\sum\limits_{i}\left(\tilde{K}_{i}^{2} + \tilde{R}_{i}^{2} \right)-\frac{2\epsilon_{\rm dd}c_{\rm d}}{\overline{\tilde{R}}^{3}} \right.\nonumber\\
& & \times\left.\left[f_{}\left(\frac{{\tilde{R}}_{x}\lambda_{x}}{{\tilde{R}}_{z}},\frac{{\tilde{R}}_{y}\lambda_{y}}{{\tilde{R}}_{z}}\right)-f_{}\left(\frac{{\tilde{K}}_{z}}{{\tilde{K}}_{x}},\frac{{\tilde{K}}_{z}}{{\tilde{K}}_{y}} \right)\right]\right\},
\label{tot_energy}
\end{eqnarray}
where the dimensionless dipolar strength $\epsilon_{\rm dd}$ is given by
\begin{equation}
\epsilon_{\rm dd} = \frac{C_{\rm dd}}{4\pi}\left(\frac{M^{3}\overline{\omega}}{\hbar^{5}} \right)^{\frac{1}{2}}N^{\frac{1}{6}}.
\label{fermi_edd}
\end{equation}
Notice that $\epsilon_{\rm dd}$ depends on the particle number and on the trap frequencies. This is in contrast with the dipolar Bose gas, where, for a system with s-wave scattering length $a_{\rm s}$, the corresponding dimensionless dipolar strength $\epsilon_{\rm dd} $ is given by
\begin{equation}
\epsilon_{\rm dd} = \frac{C_{\rm dd}M}{12\pi\hbar^{2}a_{\rm s}}.
\label{bose_edd}
\end{equation}
The difference between (\ref{fermi_edd}) and (\ref{bose_edd}) has important consequences for the behavior and the tunability of the system, as will be explained in more details below.

As a consequence of the symmetry of the anisotropy function $f(x,y)=f(y,x)$, the energy (\ref{tot_energy}) possesses the same symmetry with respect to the plane $xOy$ in both $K$- and $R$-space. On the one hand, this implies that, in a cylinder-symmetric trap, where we have $\lambda_{x}=\lambda_{y}$, the extrema of the energy satisfy ${R}_{x}={R}_{y}$. On the other hand, since the trap geometry does not influence the exchange contribution to the total energy, we conclude that the momentum distribution of a dipolar Fermi gas remains cylinder-symmetric even in the case of a triaxial trap geometry, i.e., one has ${K}_{x}={K}_{y}$, provided that the dipoles are aligned along the $Oz$ direction. Therefore, in the expression above, $f_{}({{\tilde{K}}_{z}}/{{\tilde{K}}_{x}},{{\tilde{K}}_{z}}/{{\tilde{K}}_{y}} )$ can be simplified to $f_{s}({{\tilde{K}}_{z}}/{{\tilde{K}}_{x}})$ without loss of generality. Furthermore, noticing the limits
\begin{equation}
\lim\limits_{y\rightarrow x} x\frac{\partial }{\partial x} f(x,y) = \lim\limits_{y\rightarrow x} y\frac{\partial }{\partial y} f(x,y) = -1 + \frac{\left(2\!+\!x^{2}\right)f_{s}(x)}{2(1\!-\!x^{2})},
\label{ani_f_identity}
\end{equation}
we conclude that Eqs.~(\ref{k_dynamics}) reduce to the single condition
\begin{eqnarray}
\tilde{K}_{z}^{2} - \tilde{K}_{x}^{2} & = & \epsilon_{\rm dd}C\left({\bf \tilde{R}},\tilde{K}_{x},\tilde{K}_{z}\right)
\label{dim_less_mom_def_one}
\end{eqnarray}
with the function
\begin{equation}
C\left({\bf \tilde{R}},\tilde{K}_{x},\tilde{K}_{z}\right) =  \frac{3c_{\rm d}}{\overline{\tilde{R}}^{3}}\left[\!-\!1\!+\!\frac{\left(2\tilde{K}_{x}^{2}\!+\!\tilde{K}_{z}^{2}\right)f_{s}\left({\tilde{K}_{z}}/{\tilde{K}_{x}} \right)}{ 2\left(\tilde{K}_{x}^{2} \!-\! \tilde{K}_{z}^{2} \right)} \right],
\label{dim_less_mom_def}
\end{equation}
which was already found in our previous work concerning the cylinder-symmetric trap \cite{ourpaper}. In that work, we traced Eq.~(\ref{dim_less_mom_def_one}) back to the Fock exchange term.

We emphasize that the cylindric symmetry in momentum space also holds in the dynamic case, since neither modulating the trap frequencies nor turning them off affects the symmetries of the exchange term.

The equation (\ref{tf_eqs}) for the Thomas-Fermi radius in the $i$-th direction can, thus, be written as
\begin{equation}
\frac{1}{\omega_{i}^{2}}\frac{d^{2}{\tilde{R}}_{i}}{dt^{2}} = -\tilde{R}_{i} + \sum\limits_{j}^{}\frac{\tilde{K}_{j}^{2}} {3\tilde{R}_{i}} - \epsilon_{\rm dd}Q_{i}\left({\bf \tilde{R}},\tilde{K}_{x},\tilde{K}_{z}  \right),
\label{eqn:dim_less_tf_eqs}
\end{equation}
with the corresponding simplifications in the $Q_{i}$-func\-tions (\ref{subeqn3:Q_functions}).

The equations (\ref{dim_less_part_num_conse}), (\ref{dim_less_mom_def_one}), and (\ref{eqn:dim_less_tf_eqs}) describe both the static and dynamic properties of a triaxially trapped dipolar Fermi gas in the hydrodynamic regime and represent the main result of the present article. In what follows we shall explore their solutions in different cases of interest such as the conditions for stable equilibrium, the low-lying oscillations around the equilibrium positions, and the expansion of the gas after release from the trap, i.e., the time-of-flight dynamics.

\begin{figure}
\includegraphics[scale=.85]{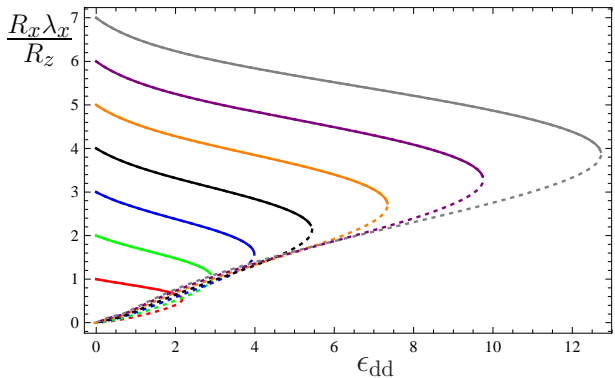}
\caption{(Color Online) Aspect ratio in real space $R_{x}\lambda_{x}/R_{z}$ for a cylinder-symmetric trap with $\lambda_{x}=\lambda_{y}=7,6,5,4,3,2,1$ (top to bottom). The upper branch (continuous) corresponds to a local minimum of the total energy, while the lower branch (dotted) represents an extremum but not a minimum.}
\label{asp_ra_real_sp_one_lambda}
\end{figure}

In terms of the dimensionless quantities introduced above, the chemical potential (\ref{mu_unitfull}) is given by
\begin{equation}
\frac{\mu}{E_{F}} = \sum\limits_{i}^{}\frac{\tilde{K}_{i}^{2}}{3}-\frac{7c_{\rm d}\epsilon_{\rm dd}}{8\overline{\tilde{R}}}\!\left[f_{}\!\left(\frac{\tilde{R}_{x}\lambda_{x}}{\tilde{R}_{z}},\frac{\tilde{R}_{y}\lambda_{y}}{\tilde{R}_{z}}\right)-f_{s}\!\left(\frac{\tilde{K}_{z}}{\tilde{K}_{x}} \right)\right].
\label{chem_pot}
\end{equation}
This expression can also be obtained through the virial theorem for dipolar gases \cite{PhysRevA.63.033606}
\begin{equation}
E_{\rm kin} - E_{\rm tr} + 3E_{\rm int}/2=0,
\label{virial}
\end{equation}
together with the scaling relation for the particle number 
\begin{equation}
N\mu = 5E_{\rm kin}/3+E_{\rm tr}+2E_{\rm int}.
\end{equation}

Now that we have explained in detail how our equations of motion arise and how they are expressed in dimensionless units, we are allowed to drop the {\LARGE ${}_{\tilde{}}$} 's and obtain a cleaner notation without any danger of misunderstandings.

\section{\label{stat_prop} Static properties}
The static properties of a dipolar Fermi gas are obtained from Eqs.~(\ref{eqn:dim_less_tf_eqs}) by requiring the left-hand side to vanish. However this only gives us the conditions for an extremal mean-field energy. Since the dipolar interaction also contains an attractive part, it is useful to have a criterion for deciding whether a given state, i.e., a point $\left({\bf {R}},{K}_{x},{K}_{z}\right)$ in the five-dimensional space of variational parameters, is stable or unstable. To that end we turn to the total energy, given by Eq.~(\ref{tot_energy}), which shall be minimized under the constraint $\overline{R}^{3}\overline{K}^{3}=1$ due to particle number conservation. A dimensional analysis of the energy (\ref{tot_energy}) shows that the system cannot have a global minimum for any non-vanishing $\epsilon_{\rm dd}$. This can be roughly seen by noticing that the stabilization comes from the factor $K_{}^{2} \sim R_{}^{-2}$ whereas the dipolar interaction goes with $R_{}^{-3}$, rendering the energy not bounded from below. Nonetheless, for weak enough interactions a local minimum might exist, to which the system would return after a small perturbation. The regions satisfying this property will be called stable, while inflection points and local maxima will be denoted unstable equilibrium points. The mathematical criterion behind this classification is given by the eigenvalues of the Hessian matrix associated with the four effectively independent variables of the problem.

\begin{figure}
\includegraphics[scale=.8]{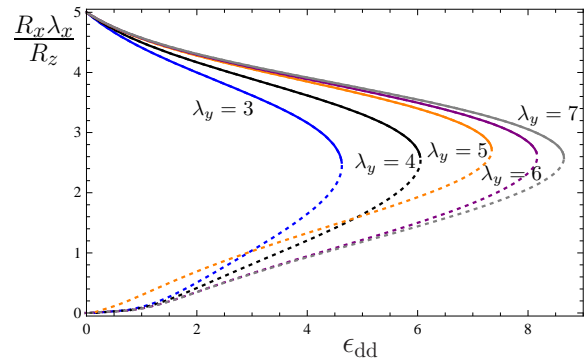}
\caption{(Color Online) Aspect ratio in real space $R_{x}\lambda_{x}/R_{z}$ for a triaxial trap with $\lambda_{x}=5$ for various values of $\lambda_{y}=3,4,5,6,7$. Notice that, for a fixed $\lambda_{x}$, making $\lambda_{y}$ larger corresponds to flattening the trap in the direction perpendicular to the dipoles, allowing for stable configurations for larger interaction strengths.}
\label{asp_ra_real_sp}
\end{figure}

One of the consequences of the unboundedness of the internal energy is that, for each value of the interaction strength $\epsilon_{\rm dd}$ for which the system presents a stable configuration, there is also another unstable one. This can be seen by considering the aspect ratio of the cloud, which is depicted in Fig.~\ref{asp_ra_real_sp_one_lambda} for different values of the trap aspect ratio $\lambda_{x}=\lambda_{y}=\lambda_{}$ as functions of $\epsilon_{\rm dd}$. Here, we recognize that the stable branch (continuous) of the real space aspect ratio starts at $\epsilon_{\rm dd}=0$ with $R_{x}=R_{z}=1$ and extends itself until the value $\epsilon_{\rm dd}^{\rm crit}$, where it meets the unstable branch (dotted). For $\epsilon_{\rm dd}>\epsilon_{\rm dd}^{\rm crit}$, no stationary solution for the equations (\ref{eqn:dim_less_tf_eqs}) exists. The unstable branch, on the other hand, possesses a vanishing aspect ratio for $\epsilon_{\rm dd}=0$. This is due to the fact that the DDI tends to stretch the sample along the polarization direction. For a small value of $\epsilon_{\rm dd}$, the unbounded energy solution is obtained with $R_{x}\rightarrow0$ and, consequently, $R_{x}/R_{z}\rightarrow 0$, although the TF radius in the axial direction $R_{z}$ remains finite. We remark that the upper branch corresponds to a local minimum of the energy such that the Hessian matrix has only positive eigenvalues, while the lower one is an extremum but not a minimum, corresponding to a Hessian matrix with at least one negative eigenvalue. The corresponding graph for a dipolar BECs in the Thomas-Fermi regime bears a crucial difference: unstable solutions only become available for $\epsilon_{\rm dd}>1$ \cite{PhysRevLett.92.250401}. The physical reason for this effect is that in dipolar BECs the stabilization comes from the contact interaction, which scales with $R^{-3}$, just like the DDI.

In order to study the effect of a triaxial trap on the static properties of a dipolar Fermi gas, we explore further the symmetry $f(x,y)=f(y,x)$ of the anisotropy function as defined by Eq.~(\ref{aniso_func}). Due to this symmetry, we only need to discuss the aspect ratio $R_{x}\lambda_{x}/R_{z}$ since the properties of $R_{y}\lambda_{y}/R_{z}$ can be obtained by analogy. As indicated in Fig.~\ref{asp_ra_real_sp}, varying $\lambda_{y}$ for fixed $\lambda_{x}$ clearly affects the stability of the system. For $\lambda_{y}>\lambda_{x}$, stable solutions are admitted for larger values of $\epsilon_{\rm dd}$, i.e., $\epsilon_{\rm dd}^{\rm crit}$ is shifted to the right, whereas in the case $\lambda_{y}<\lambda_{x}$, $\epsilon_{\rm dd}^{\rm crit}$ decreases. This reflects the fact that more oblate traps tend to allow for larger $\epsilon_{\rm dd}$ because they favor the repelling part of the interaction. Another worth remarking feature in Fig.~\ref{asp_ra_real_sp} is that reducing $\lambda_{y}$ for fixed $\lambda_{x}$ reduces the value of $\epsilon_{\rm dd}^{\rm crit}$ much more than it is enlarged by increasing $\lambda_{y}$.

\begin{figure}
\includegraphics[scale=.9]{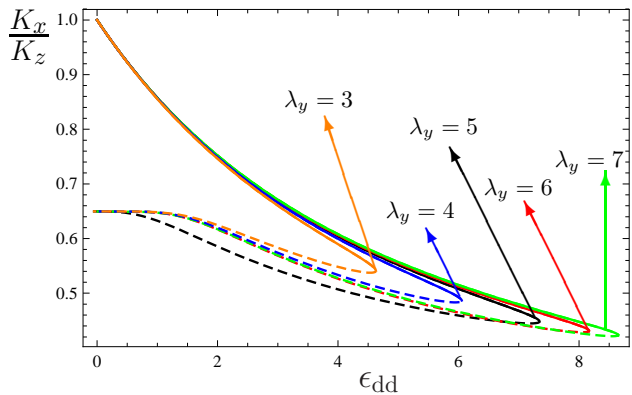}
\caption{(Color Online) Aspect ratio in momentum space $K_{x}/K_{z}$ for $\lambda_{x}=5$ and $\lambda_{y}=3,4,5,6,7$. The upper branch (continuous) corresponds to a local minimum of the total energy, while the lower branch (dashed) represents an extremum but not a minimum.}
\label{kr_asp_ra_real_sp_one_lambda}
\end{figure}

Concerning the aspect ratio in momentum space, we have studied its dependence on the dipolar strength $\epsilon_{\rm dd}$ and found an analogous behavior to the one in real space. This goes back to the property of the function $f_{s}(x)$ of changing sign at $x=1$, so that the minus sign in front of $f_{s}({{K}_{z}}/{{K}_{x}} )$ in the total energy partially compensates its dependence on the reciprocal momentum aspect ratio ${{K}_{z}}/{{K}_{x}}$, and its behavior with respect to $\epsilon_{\rm dd}$ turns out to be analogous to one in real space. This is explicitly shown in Fig.~\ref{kr_asp_ra_real_sp_one_lambda}, where the aspect ratio in momentum space $K_{x}/K_{z}$ is plotted as a function of $\epsilon_{\rm dd}$  for $\lambda_{x}=5$ and $\lambda_{y}=3,4,5,6,{\textrm{ and }}7$. The main difference, which appears in momentum space, is the observation that the unstable solution converges to a finite value of the aspect ratio as the interaction strength $\epsilon_{\rm dd}$ approaches zero. This reflects the fact that the collapse is a real-space phenomenum which is dominated by the shrinking of the radial Thomas-Fermi radius $R_{x}$, while the axial Thomas-Fermi radius $R_{z}$ remains finite. As the momentum-space variables are accounted for only by the constraint (\ref{dim_less_part_num_conse}) and the condition for momentum deformation (\ref{dim_less_mom_def_one}), both $K_{x}$ and $K_{z}$ diverge as $\epsilon_{\rm dd}$ approaches zero in the unstable branch, but their ratio always is finite.

\begin{figure}
\includegraphics[scale=.9]{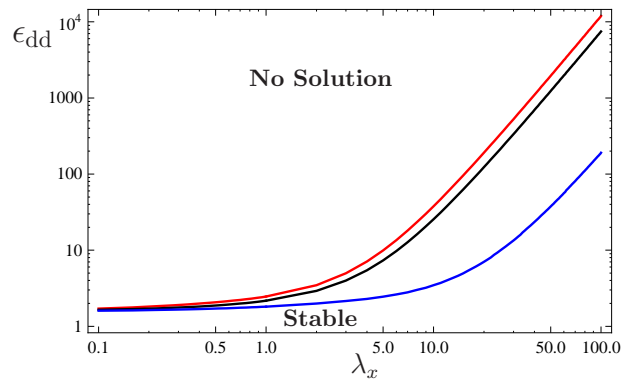}
\caption{(Color Online) Stability diagram of a dipolar Fermi gas. The middle, black curve represents the cylinder-symmetric case $\lambda_{y}=\lambda_{x}$, while the upper, red one represents the case $\lambda_{y}=5\lambda_{x}$ and the lower, blue curve is for $\lambda_{y}=\lambda_{x}/5$.}
\label{stability}
\end{figure}

To conclude our investigation of the static properties of a trapped dipolar Fermi gas, we have calculated the stability diagram for the cylinder-symmetric case $\lambda_{x}=\lambda_{y}$, where we obtain similar quantitative results as in Ref.~\cite{miyakawa:061603}, and for $\lambda_{x}\neq\lambda_{y}$, where the lack of axial symmetry has a considerable influence. The results are presented in a log-log plot in Fig.~\ref{stability}. If we consider a situation in which $\lambda_{y}=5\lambda_{x}$ (red, upper curve), we do not obtain a large variation with respect to the cylinder-symmetric case $\lambda_{y}=\lambda_{x}$ (black, middle curve). On the contrary, if we take $\lambda_{y}=\lambda_{x}/5$, appreciable differences can be noticed as $\lambda_{x}$ increases. This can be understood if one realizes that it is the weaker trap frequency which determines the highest value of $\epsilon_{\rm dd}$, for which the system remains stable. Therefore, by enlarging $\lambda_{y}$ with respect to $\lambda_{x}$ one obtains a smaller difference with respect to the case $\lambda_{y}=\lambda_{x}$ than by reducing it, explaining the effect already anticipated in Fig.~\ref{asp_ra_real_sp}. Also remarkable is the fact that, for small $\lambda$'s, the three curves lie very close to each other.

\section{\label{ll_excit} Low-lying excitations}

The low-lying excitations of a dipolar Fermi gas are studied in this section by linearizing the equations of motion around the equilibrium. This is done by assuming that at time $t$ the following ansatz is valid
\begin{equation}
R_{i}(t) = R_{i}(0) + \eta_{i}e^{i\Omega t};\qquad K_{i}(t) = K_{i}(0) + \zeta_{i}e^{i\Omega t},
\end{equation}
where $\eta_{i}$ and $\zeta_{i}$ denote the small amplitudes in real and momentum space, respectively, while $\Omega$ represents the frequency of the oscillations. Due to the cylinder symmetry in momentum space, there are only two independent momentum-space amplitudes $\zeta_{x}$ and $\zeta_{z}$, while three independent real-space amplitudes occur for a general three-dimensional trap.

\begin{figure}
\includegraphics[scale=.8]{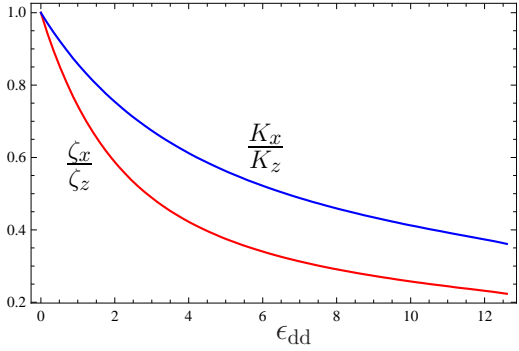}
\caption{(Color Online) The lower (red) curve shows the ratio of the amplitudes $\zeta_{x}/\zeta_{z}$ as a function of $\epsilon_{\rm dd}$ 
for $\lambda_{x}=\lambda_{y}=7$. For comparison, the stable branch of the equilibrium aspect ratio in momentum space against $\epsilon_{\rm dd}$ for $\lambda_{x}=\lambda_{y}=7$ is depicted by the upper (blue) curve.}
\label{mom_osc}
\end{figure}

Before we can derive a matrix equation for the amplitudes in real space, we must obtain the $\zeta_{i}$'s as functions of the $\eta_{i}$'s. To that end, we expand Eqs.~(\ref{dim_less_part_num_conse}) and (\ref{dim_less_mom_def_one}) up to first order in the amplitudes and get
\begin{equation}
\zeta_{i} = \left(\sum\limits_{j}\frac{\eta_{j}}{R_{j}}\right)K_{i} W_{i}
\end{equation}
with the abbreviations
\begin{eqnarray}
W_{x} & = &-\frac{K_{x}^{2}+K_{z}^{2}-\epsilon_{\rm dd}K_{z}C_{,K_{z}}}{4K_{z}^{2} + 2K_{x}^{2}-3\epsilon_{\rm dd}K_{z}C_{,K_{z}}},\nonumber\\
W_{z} & = &-\frac{2K_{z}^{2}-\epsilon_{\rm dd}K_{z}C_{,K_{z}}}{4K_{z}^{2} + 2K_{x}^{2}-3\epsilon_{\rm dd}K_{z}C_{,K_{z}}}.
\end{eqnarray}
To make the notation more succint, we have introduced here the shorthand $A_{,K_{z}}=\partial A\left({\bf {R}},{K}_{x},K_{z}\right)/\partial K_{z}$ to denote a partial derivative of the quantity $A\left({\bf {R}},{K}_{x},K_{z}\right)$ with respect to $K_{z}$ evaluated at equilibrium. These results show that the presence of the dipolar exchange term drives the momentum oscillations {\sl anisotropic}. In Fig.~\ref{mom_osc} we plot the ratio $\zeta_{x}/\zeta_{z}$ as a function of $\epsilon_{\rm dd}$ 
for $\lambda_{x}=\lambda_{y}=7$. We show also the stable branch of the corresponding equilibrium aspect ratio in momentum space, represented by the upper (blue) curve. In order to appreciate the meaning of this curve, let us consider a typical experimental situation with $N\approx4\cdot10^{4}$ KRb molecules and trap frequencies of $(\omega_{x},\omega_{y},\omega_{z})=2\pi~(40,40,280)$ Hz. By using an external electric field and tuning the electric dipole moment to $d=0.2$ Debye, one obtains the dipole-interaction strength $\epsilon_{\rm dd}\approx0.43$ which leads to an oscillation anisotropy of $\zeta_{x}/\zeta_{z}\approx0.87$ and an equilibrium momentum deformation of $K_{x}/K_{z}\approx0.93$. More striking effects result for a stronger interaction. Considering an electric dipole moment of $d=0.57$ Debye yields a larger value $\epsilon_{\rm dd}\approx3.53$ and, therefore, also larger anisotropies for both the momentum oscillation $\zeta_{x}/\zeta_{z}\approx0.45$ and equilibrium momentum distribution $K_{x}/K_{z}\approx0.64$. These results exhibit clearly the effects of the exchange term on the low-lying oscillations and make room for a clear detection of the DDI in ultracold degenerate Fermi gases.

Linearizing the equations for the TF radii (\ref{eqn:dim_less_tf_eqs}) we obtain
\begin{equation}
\left[\frac{\Omega^{2}}{\omega_{i}^{2}}\!-\!1\!-\!\frac{2K_{x}^{2}+K_{z}^{2}}{3R_{i}^{2}}\right]\eta_{i} -  \sum\limits_{j}\left[\frac{P}{R_{i}R_{j}} + \epsilon_{\rm dd}Q_{i,R_{j}}\right]\eta_{j}=0,
\label{eigen_matrix}
\end{equation}
where we have introduced the shorthand
\begin{equation}
P=\frac{2}{3}\frac{K_{x}^{4}+K_{z}^{4}+4K_{x}^{2}K_{z}^{2}-\left(2K_{x}^{2}+K_{z}^{2}\right)\epsilon_{\rm dd }K_{z}C_{,K_{z}}}{4K_{z}^{2} +2K_{x}^{2}-3\epsilon_{\rm dd}K_{z}C_{,K_{z}}},
\end{equation}
which approaches the value $2/3$, as $\epsilon_{\rm dd}$ goes to zero.

With this the study of the low-lying oscillations in a dipolar Fermi gas has been reduced to the eigenvalue problem (\ref{eigen_matrix}): the oscillation frequencies $\Omega$ are given by the square root of the corresponding eigenvalues and the eigenmodes describe the real-space motion during the oscillations.

For a non-interacting Fermi gas, this formalism recovers the oscillation frequencies of a triaxial trap as the solutions of the equation
\begin{eqnarray}
\!\!\!\!\!\!\!\!\!\!\!\!\!\!\!\!\!\!\!\!\!\!\!&&3{\Omega^{(0)}}^{6}-8{\Omega^{(0)}}^{4}\!\!\left(\omega_{x}^{2}+\omega_{y}^{2}+\omega_{z}^{2}\right)+20{\Omega^{(0)}}^{2}\!\!\left(\omega_{x}^{2}\omega_{y}^{2}+\omega_{x}^{2}\omega_{z}^{2}+\right. \nonumber\\
\!\!\!\!\!\!\!\!\!&&\left.\omega_{y}^{2}\omega_{z}^{2}\right)-48\omega_{x}^{2}\omega_{y}^{2}\omega_{z}^{2}=0.
\label{non_int_freqs}
\end{eqnarray}

This result is in agreement with Ref.~\cite{PhysRevA.63.013606}, where a deeper analysis, initially devised for BECs \cite{PhysRevA.56.R2533}, is carried out. It is shown there that, despite the lack of an obvious spatial symmetry, the wave equation for the hydrodynamic modes is separable in elliptical coordinates. We remark that the solutions of Eq. (\ref{non_int_freqs}) reduce to the respective frequencies in the presence of cylindrical \cite{amoruso} or spherical symmetries \cite{PhysRevLett.83.5415}, where this problem was first tackled.

In the following, we discuss separately the effects of the DDI in cylindric and triaxial traps. The modification introduced in Eq.~(\ref{non_int_freqs}) due to the inclusion of the DDI makes this equation too cumbersome to be displayed here. The same is true for the corresponding solutions. For this reason, we shall provide detailed expressions for the three oscillation frequencies only in the case of cylinder symmetry.

\subsection{Oscillation frequencies in cylinder-symmetric traps}

In the presence of cylinder symmetry, we find three well characterized oscillation modes: one two-dimensional mode, the radial quadrupole, and two three-dimensional ones, the monopole and the quadrupole modes.

\begin{figure}
\includegraphics[scale=.9]{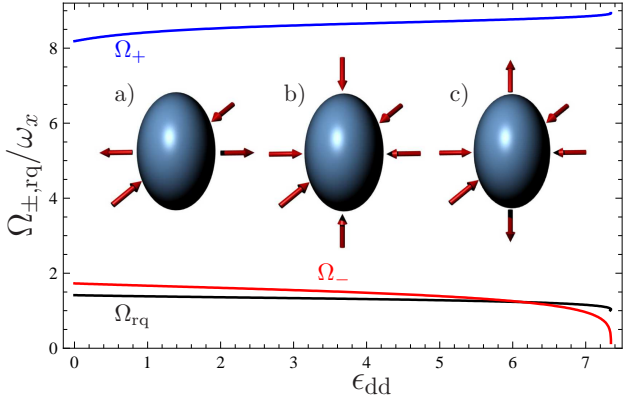}
\caption{(Color Online) Oscillation modes in the cylinder-symmetric configuration for $\lambda=5$. The oscillation frequencies of the radial quadrupole $\Omega_{\rm rq}$ (black), the monopole $\Omega_{+}$ (blue, dark grey), and the three-dimensional quadrupole mode $\Omega_{-}$ (red, light grey) are plotted as functions of the dipolar interaction strength $\epsilon_{\rm dd}$ in units of $\omega_{x}$. The inset shows the behavior of the corresponding eigenmodes. Part a) characterizes the radial quadrupole eigenmode, while b) and c) refer to the monopole and quadrupole mode, respectively.}
\label{oscill}
\end{figure}

The first mode we consider is the radial quadrupole mode, depicted in Fig.~\ref{oscill}a). It is characterized by a vanishing amplitude in the $Oz$-direction, while the oscillations in $Ox$- and $Oy$-directions have the same amplitude but are completely out-of-phase. We find the frequency $\Omega^{}_{\rm rq}$ to be given by
\begin{eqnarray}
\!\!\!\!\!\!\!\Omega_{\rm rq} & = &\omega_{x}\left\{2+\frac{3c_{\rm d}\epsilon_{\rm dd}}{R_{x}^{4}R_{z}^{}}\frac{R_{x}^{2}\lambda^{2}}{R_{z}^{2}}\right.\nonumber\\
\!\!\!\!\!& &\!\!\!\!\!\!\!\!\!\!\!\!\!\!\!\!\!\!\!\!\!\!\!\!\times\left.\frac{2\left( R_{z}^{2}-R_{x}^{2}\lambda^{2}\right)-\left( 4R_{z}^{2}+R_{x}^{2}\lambda^{2}\right)f_{s}\left(R_{x}\lambda/R_{z}\right)}{\left(R_{z}^{2}-R_{x}^{2}\lambda^{2} \right)^{2}} \right\}^{1/2}\!\!\!\!\!\!\!\!,
\label{2d_quadr}
\end{eqnarray}
where the TF radii $R_{x}$ and $R_{z}$ correspond to the static values calculated in Section \ref{stat_prop}.

\begin{figure}
\includegraphics[scale=.9]{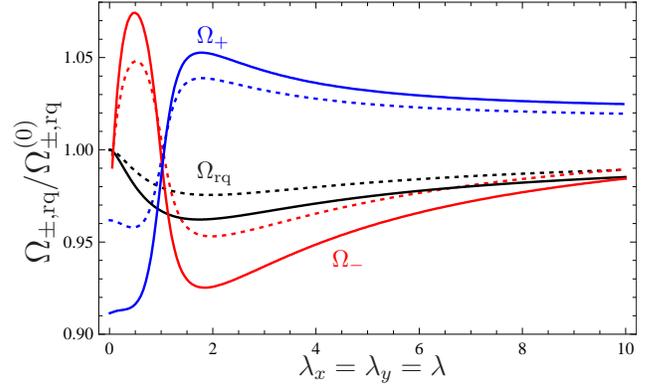}
\caption{(Color Online) Low-lying oscillation frequencies in units of the corresponding non-interacting values as functions of $\lambda$ for $\epsilon_{\rm dd}=0.8$ (dotted) and $\epsilon_{\rm dd}=1.2$ (continuous). The monopole and quadrupole modes are displayed in blue (dark grey) and red (light grey), respectively, while the radial quadrupole mode is shown in black.}
\label{oscill_Lambda}
\end{figure}

The radial quadrupole mode can be experimentally excited by adiabatically deforming the circular trap in the $xOy$-plane into an ellipse and suddenly switching off the deformation. In the case of a Fermi gas with contact interaction, the radial quadrupole mode was used to probe the transition from the collisionless to the hydrodynamic regime throughout the BEC-BECS-crossover \cite{altmeyer:033610}. For a dipolar Fermi gas, a similar experiment could be thought of, where controlling the collisions through applied electric fields would play the role of a Feshbach resonance to tune the system all the way from a ballistic to a hydrodynamic behavior in the normal phase. Fig.~\ref{oscill} depicts $\Omega^{}_{\rm rq}$ as a function of $\epsilon_{\rm dd}$ for $\lambda=5$ in units of $\omega_{x}$. We find that, for a given $\lambda$, the frequency $\Omega^{}_{\rm rq}$ is quite insensitive to changes in the interaction over the range of values in which the gas is stable. Although we have varied the trap anisotropy $\lambda$ from $0.2$ up to $20$, no significant alteration of this behavior could be detected. Fig.~\ref{oscill_Lambda} shows the dependence of $\Omega^{}_{\rm rq}$ on $\lambda$ for $\epsilon_{\rm dd}=0.8$ and $\epsilon_{\rm dd}=1.2$ in units of its non-interacting value, i.e., $\Omega^{(0)}_{\rm rq}=\sqrt{2}\omega_{x}$, which is directly given in Eq.~(\ref{2d_quadr}) by setting $\epsilon_{\rm dd}=0$. We remark that the function in the second line of Eq.~(\ref{2d_quadr}) is a function of the ratio $R_{x}\lambda/R_{z}$ alone which approaches the value $-16/35$ as $R_{x}\lambda/R_{z}$ tends to $1$, so that no divergence arises for $R_{z}=R_{x}\lambda$. We would also like to point out that, despite the fact that the radial quadrupole mode is inherent to cylinder-symmetric systems, its calculation requires that one starts from a triaxial framework, which is then specialized to axial symmetry. This is the reason, why this important mode was not explored in initial studies of dipolar Fermi gases in the hydrodynamic \cite{ourpaper} or collisionless regime \cite{1367-2630-11-5-055017}.

We now concentrate on the three-dimensional mon\-o\-pole and quadrupole modes. The first, also known as breathing mode, is a compression mode characterized by an in-phase oscillation in all three directions and is denoted with an index $+$. The second, in analogy with the radial quadrupole mode, is an out-of phase oscillation in radial and axial directions and is denoted with an index $-$. In a spherical trap, these modes are decoupled from each other, but, here, they are coupled due to the cylinder symmetry of the trap. Their frequencies are given by
\begin{equation}
\Omega^{}_{\pm} = \frac{\omega_{x}^{}}{\sqrt{2}}\sqrt{M_{xx}+M_{zz}\pm\sqrt{2M_{xz}^{2}+\left(M_{xx}-M_{zz}\right)^{2}}},
\end{equation}
together with the abbreviations
\begin{eqnarray}
M_{xx} & = & 2+\frac{2P}{R_{x}^{2}}+\frac{c_{\rm d}\epsilon_{\rm dd}}{R_{x}^{4}R_{z}^{}}\!\!\left[\frac{ \!-2R_{z}^{4}\!+\!7R_{z}^{2}R_{x}^{2}\lambda^{2}\!-\!5R_{x}^{4}\lambda^{4}}{\left(R_{z}^{2}-R_{x}^{2}\lambda^{2} \right)^{2}}\right.\nonumber\\
\!\!\!\!\!\!\!\!\!\!\!\!& &\!\!\!\!\!\!\!\!\!\!\!\!\!\!\!\!\!\! \left.-\frac{3R_{x}^{2}\lambda^{2}\left( 2R_{z}^{2}\!+\!3R_{x}^{2}\lambda^{2}\right)}{2\left(R_{z}^{2}-R_{x}^{2}\lambda^{2} \right)^{2}}f_{s}\left(\frac{R_{x}\lambda}{R_{z}}\right)+2f_{s}\left(\frac{K_{z}}{K_{x}} \right)\right] ,\nonumber\\
\frac{M_{zz}}{\lambda^{2}} & = & \left.2+\frac{P}{R_{z}^{2}}+\frac{c_{\rm d}\epsilon_{\rm dd}}{R_{x}^{2}R_{z}^{3}}\!\!\left[\frac{2\left( 4R_{z}^{4}-5R_{z}^{2}R_{x}^{2}\lambda^{2}+R_{x}^{4}\lambda^{4}\right)}{\left(R_{z}^{2}-R_{x}^{2}\lambda^{2} \right)^{2}}\right.\right.\nonumber\\
& &\!\!\!\!\!\!\!\!\!\!\!\!\!\!\!\left.\left.-\frac{3R_{z}^{2}\left( 3R_{z}^{2}+2R_{x}^{2}\lambda^{2}\right)}{\left(R_{z}^{2}-R_{x}^{2}\lambda^{2} \right)^{2}}f_{s}\left(\frac{R_{x}\lambda}{R_{z}}\right) +f_{s}\left(\frac{K_{z}}{K_{x}} \right)\right]\right.,\nonumber\\
\frac{M_{xz}}{2\lambda} & = & \left.\frac{P}{R_{x}R_{z}}+\frac{c_{\rm d}\epsilon_{\rm dd}}{R_{x}^{3}R_{z}^{2}}\!\!\left[-\frac{ R_{z}^{4}+R_{z}^{2}R_{x}^{2}\lambda^{2}-2R_{x}^{4}\lambda^{4}}{\left(R_{z}^{2}-R_{x}^{2}\lambda^{2} \right)^{2}}\right.\right.\nonumber\\
& &\!\!\!\!\!\!\!\!\!\!\left.\left.+\frac{15R_{z}^{2}R_{x}^{2}\lambda^{2}}{2\left(R_{z}^{2}-R_{x}^{2}\lambda^{2} \right)^{2}}f_{s}\left(\frac{R_{x}\lambda}{R_{z}}\right) +f_{s}\left(\frac{K_{z}}{K_{x}} \right)\right]\right..\nonumber
\end{eqnarray}

The dependence of the mono- and quadrupole oscillation frequencies on the DDI strength $\epsilon_{\rm dd}$ for a fixed trap anisotropy $\lambda$ is shown in Fig.~\ref{oscill}. We find that the frequencies behave for different values of $\lambda>1$ qualitatively like in Fig.~\ref{oscill}, where we have $\lambda=5$: The monopole frequency increases monotonically and its derivative with respect to $\epsilon_{\rm dd}$ blows up as $\epsilon_{\rm dd}^{\rm crit}$ is 
approached. On the contrary, the frequency of the other two modes decrease and their inclinations fall down abruptly in the neighborhood of $\epsilon_{\rm dd}^{\rm crit}$. In the case of $\lambda<1$, the monopole frequency changes its behavior, which ceases to be monotonic in $\epsilon_{\rm dd}$. It grows for small $\epsilon_{\rm dd}$, but, as the critical interaction strength is approached, it starts decreasing as $\epsilon_{\rm dd}$ grows. The radial quadrupole and three-dimensional quadrupole frequencies behave as functions of $\epsilon_{\rm dd}$ for $\lambda<1$ qualitatively nearly the same as for $\lambda>1$. For specific values of $\lambda$, though, they might cease to be monotonically decreasing. The characteristic feature here is that, for both $\lambda<1$ and $\lambda>1$, the three-dimensional quadrupole frequency vanishes at $\epsilon_{\rm dd}^{\rm crit}$, as a signal of global collapse of the gas.

\begin{figure}
\includegraphics[scale=.7]{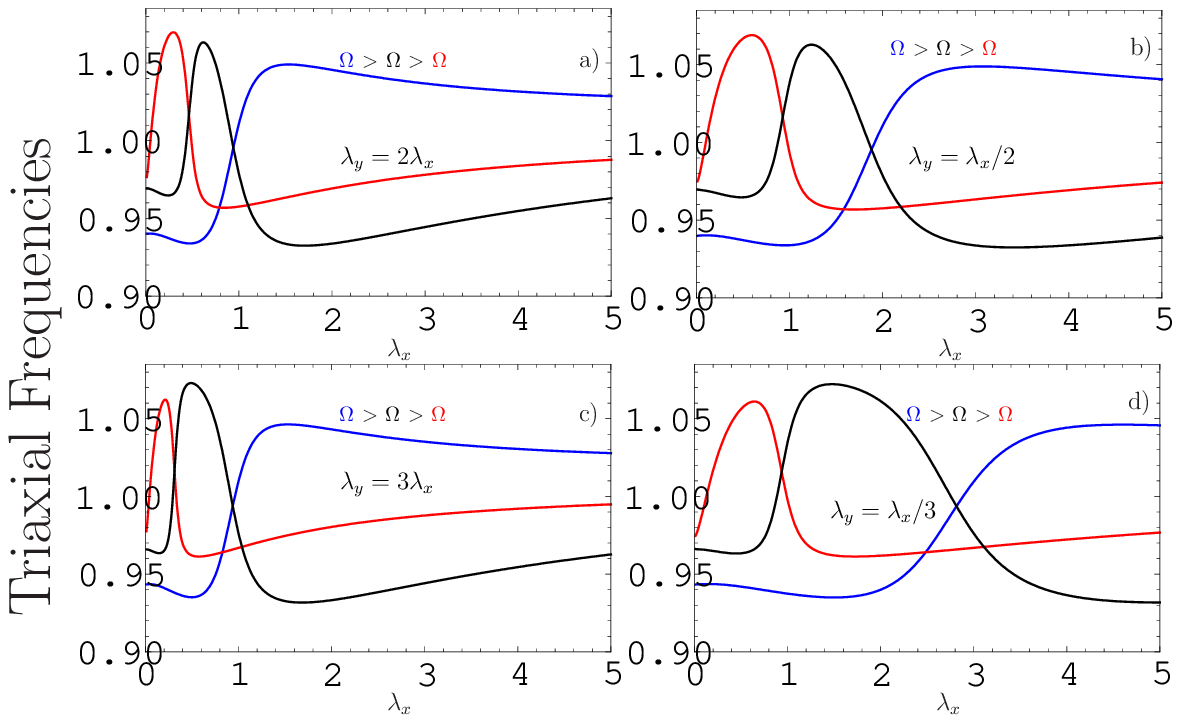}
\vspace{.25cm}

\includegraphics[scale=.8]{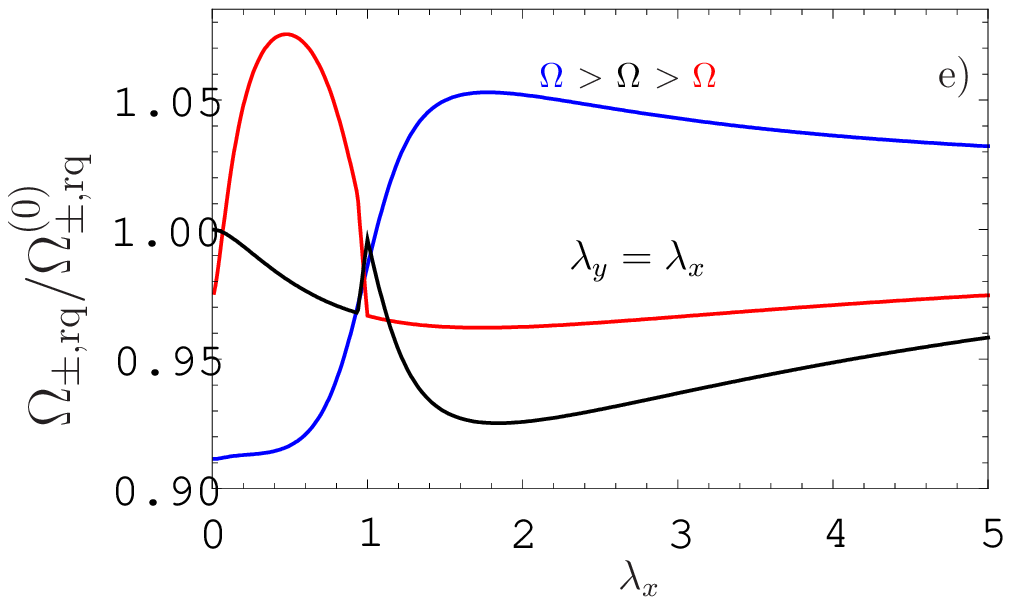}
\caption{(Color Online) Low-lying oscillation frequencies for triaxial traps as functions of the trap aspect ratio $\lambda_{x}$ for $\epsilon_{\rm dd}=1.2$ and different values of $\lambda_{y}/\lambda_{x}$. The frequencies are normalized by their respective non-interacting values. The curves are marked with the colors blue (dark grey), black, and red (light grey) corresponding to the highest, moderate, and lowest value, respectively. Figs.~\ref{oscill_Lambda_triaxial} a-d) show the mixing of the two quadrupole-like modes, which go continuously over into each other. In Fig.~\ref{oscill_Lambda_triaxial} e) a level-crossing in the cylinder-symmetric case becomes apparent through an abrupt permutation of the quadrupole modes (see Fig.~\ref{level_crossing} for more details).}
\label{oscill_Lambda_triaxial}
\end{figure}

How the oscillation frequencies depend on $\lambda$ is shown in Fig.~\ref{oscill_Lambda} for $\epsilon_{\rm dd}=0.8$ (dotted) and $\epsilon_{\rm dd}=1.2$ (continuous). For $\lambda<1$, the quadrupole frequency is larger in comparison to the non-interacting case, while the contrary is true for the monopole frequency. As $\lambda$ eventually becomes larger than $1$, the monopole (quadrupole) becomes larger (smaller) than in the absence of interactions. Concerning the radial quadrupole frequency, it turns out to be the most insensitive with respect to the dipolar interaction and is always smaller in the presence of the DDI with a minimum around $\lambda\approx1.74$. The behavior of the three-dimensional modes normalized by their non-interacting values agrees qualitatively with both dipolar BECs \cite{PhysRevLett.92.250401} and with dipolar Fermi gases in the collisionless regime \cite{1367-2630-11-5-055017}.

\subsection{Oscillation frequencies in triaxial traps}

In the most general case, i.e., in absence of cylinder symmetry, the oscillation modes do not to behave as indicated in the inset of the Fig.~\ref{oscill}: each of the three modes becomes a superposition of in- and out-of-phase oscillations in all three spatial directions. For this reason, the modes are better characterized by their frequencies and these are naturally mixed, even if one looks at the cylinder-symmetric limit of the triaxial solutions. We thus plot the frequencies according to their values and the colors blue (dark grey), black, and red (light grey) correspond to the highest, moderate, and lowest value, respectively. We exhibit in Fig.~\ref{oscill_Lambda_triaxial} the dependence of these frequencies on $\lambda_{x}$ for different values of $\lambda_{y}/\lambda_{x}$ and $\epsilon_{\rm dd}=1.2$, with the frequencies normalized by their respective non-interacting values. The situations $\lambda_{y}/\lambda_{x}=2,1/2,3,1/3$ correspond to Figs.~\ref{oscill_Lambda_triaxial} a), b), c) and d), respectively. These pictures show explicitly that the two quadrupole-like modes, denoted by the colors red (light grey) and black, are now mixed. If the cylinder-symmetric situation is considered, a level-crossing becomes evident at $\lambda_{}\approx 0.94$, shown in Fig.~\ref{oscill_Lambda_triaxial} e). In contrast to the bosonic case \cite{PhysRevA.75.015604}, the DDI affects the value of $\lambda$ at which the level-crossing takes place. For this reason, instead of a discontinuous transition, as for dipolar bosons, there is a steep continuous line in both the radial and the three-dimensional quadrupole modes for the trap anisotropy range $0.94<\lambda_{}<1$.

\begin{figure}
\includegraphics[scale=.9]{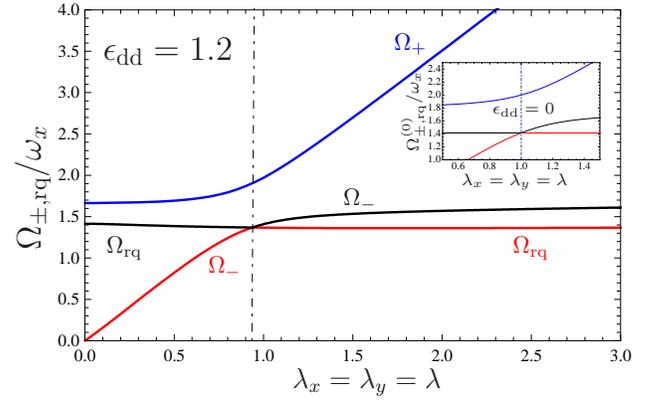}
\caption{(Color Online) Frequencies of the low-lying oscillations in units of $\omega_{x}$ in the cylinder-symmetric configuration as functions 
of $\lambda$ for $\epsilon_{\rm dd}=1.2$. The colors blue (dark grey), black, and red (light grey) label the frequencies in decreasing value. At
 $\lambda\approx0.94$ a level-crossing takes place between the two quadrupole modes. In the absence of the interaction,
the level-crossing happens precisely at $\lambda=1$, as can be seen in the inset. This difference explains the steep curves in Fig.~\ref{oscill_Lambda_triaxial}e) for $0.94<\lambda<1$.}
\label{level_crossing}
\end{figure}

The level-crossing in the cylinder-symmetric case can be seen more clearly in Fig.~\ref{level_crossing}, where the frequencies are plotted in units of $\omega_{x}$ for $\epsilon_{\rm dd}=1.2$. The vertical line marks the level-crossing, which takes place at $\lambda_{}\approx 0.94$. The inset contains a zoomed picture of the frequencies for $\epsilon_{\rm dd}=0$. There,  the level-crossing happens at $\lambda_{}= 1$, explaining the steep lines which show up in the spectra of the quadrupole modes in Fig.~\ref{oscill_Lambda_triaxial} e). This shift of the level-crossing can be traced back to the Fock exchange interaction, which is absent in dipolar Bose-Einstein condensates.

\section{\label{tof_exp} TOF expansion}
Time-of-flight expansion experiments are a key diagnostic tool in the field of ultracold quantum gases. In BEC's, for example, the effects of the magnetic DDI were observed for the first time in $^{52}$Cr by measuring the time dependence of the aspect ratios for two different polarization directions after release from a triaxial trap \cite{PhysRevLett.95.150406}. In this section, we explore the corresponding problem for a strong dipolar normal Fermi gas.

Dipolar effects are expected to be observed in polar molecules, due to their large electric dipole moment. Trapping and cooling these molecules requires a strong confinement in the polarization axes, to assure robustness against collapse. Therefore, the suppression of the attractive part of the DDI indicates that this system is better described by normal hydrodynamics. Initial estimates of the relaxation time for polar molecules suggest that this reasoning remains valid during the whole TOF expansion \cite{ourpaper}. Thus, we expect the dynamics of the dipolar Fermi gas to be described by the equations
\begin{equation}
\frac{1}{\omega_{i}^{2}}\frac{d^{2}{{R}}_{i}}{dt^{2}} = \sum\limits_{j}^{}\frac{{K}_{j}^{2}} {3{R}_{i}} - \epsilon_{\rm dd}Q_{i}\left({\bf {R}},{K}_{x},K_{z}  \right),
\label{eqn:dim_less_tf_eqs_tof}
\end{equation}
together with the conditions for number conservation (\ref{dim_less_part_num_conse}) and momentum deformation (\ref{dim_less_mom_def}). Notice that equation (\ref{eqn:dim_less_tf_eqs_tof}) differs from (\ref{eqn:dim_less_tf_eqs}) only due to the absent term $-R_{i}$, which is responsible for the trapping potential. In the following, we discuss the results obtained by solving these equation numerically, using the static values of Section \ref{stat_prop} for the initial conditions of the parameters $R_{i}(0)$ and $K_{i}(0)$ as well as $\dot{R}_{i}(0)=0$ and $\dot{K}_{i}(0)=0$.

Until now, only axial symmetric traps were involved in experimental investigations of dipolar Fermi gases. Nevertheless, we have learned from studies of dipolar BECs how useful triaxial traps can be, for instance in the context of time-of-flight experiments.

\begin{figure}
\includegraphics[scale=.75]{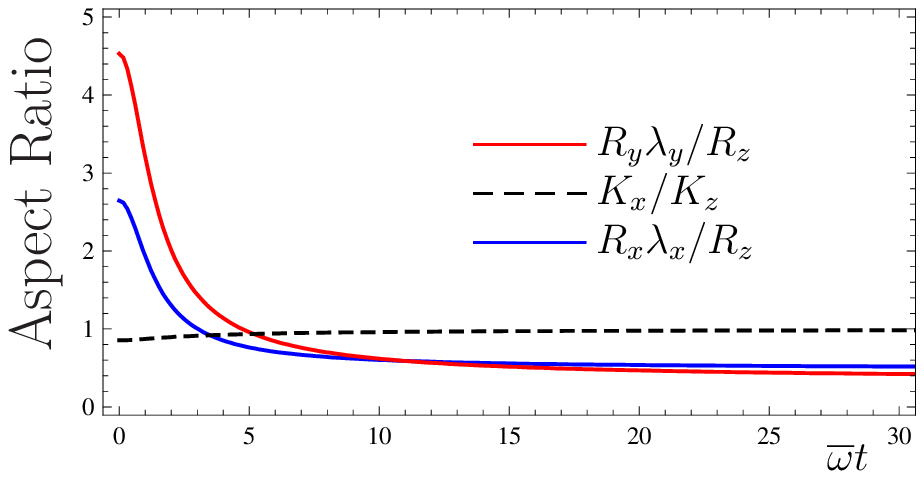}

\includegraphics[scale=.75]{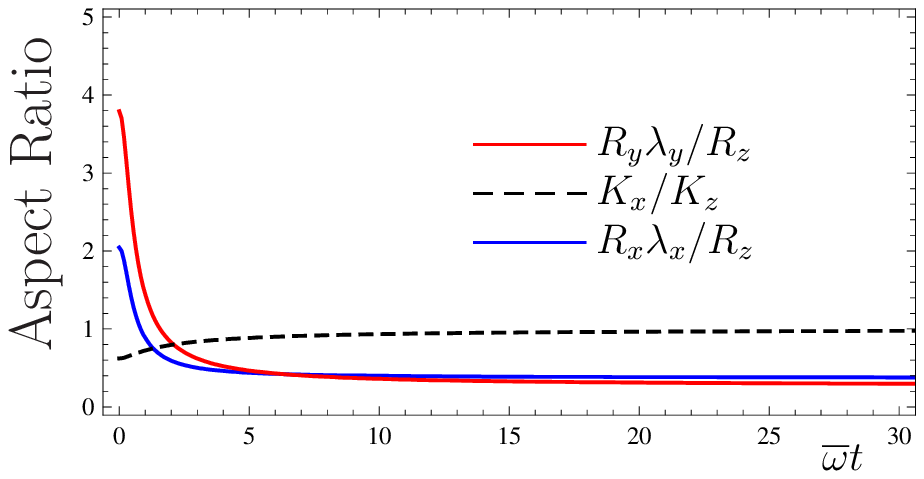}
\caption{(Color Online) Aspect ratios in real and momentum space as functions of time. The dashed curve corresponds to the aspect ratio in momentum space $K_{x}/K_{z}$ while the red (light grey) and blue (dark grey) curves correspond to the real-space aspect ratios $R_{y}\lambda_{y}/R_{z}$ and $R_{x}\lambda_{x}/R_{z}$, respectively. The trap is characterized by $\lambda_{x}=3$ and $\lambda_{y}=5$. The top curves correspond to $\epsilon_{\rm dd}=1 $ while the bottom ones are for $\epsilon_{\rm dd}= 3.5$.}
\label{tof1}
\end{figure}

Concerning the momentum space, we obtain in the triaxial case similar results as for the cylinder-symmetric one \cite{ourpaper}, where the aspect ratio $K_{x}/K_{z}$ becomes asymptotically unity as a result of local equilibrium in the absence of the trap. The anisotropic aspect ratios $R_{x}\lambda_{x}/R_{z}$ and $R_{y}\lambda_{y}/R_{z}$ are plotted as functions of time in Fig.~\ref{tof1} for $\lambda_{x}=3$ and $\lambda_{y}=5$. In the upper and lower graphs we have set $\epsilon_{\rm dd}=1$ and $\epsilon_{\rm dd}=3.5$, respectively, and we find that both aspect ratios become smaller than $1$ in the course of time. Also for traps with $\lambda_{x}<1$ or $\lambda_{y}<1$, an inversion of the corresponding aspect ratio takes place, but in the opposite direction. Such an inversion is typical for the hydrodynamic regime and was already observed for a two-component, normal Fermi gas with strong contact interaction \cite{ohara,PhysRevLett.91.020402}.

The value $\epsilon_{\rm dd}=3.5$, chosen above, corresponds to $N=4\cdot10^{4}$ KRb-molecules, with a dipole moment of $d\approx0.51$ Debye induced by an applied electric field and with trap frequencies characterized by $\omega_{z}=2\pi\times280$, $\lambda_{x}=3$, and $\lambda_{y}=5$. Simple arguments like the ones given in Ref.~\cite{ourpaper}, show that the hydrodynamic character of the expansion holds, at least, for $\overline{\omega} t\ll 42$. Given that the trapping frequencies can be changed at will over a wide range, the prospects for observing hydrodynamic expansion in dipolar gases of heteronuclear molecules out of triaxial traps are quite promising.

\begin{figure}[t]
\includegraphics[scale=.9]{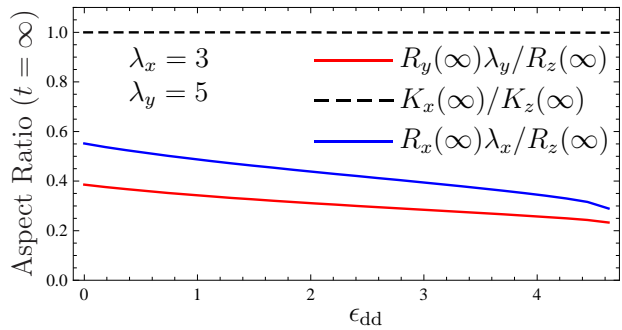}
\caption{(Color Online) Asymptotic behavior in time of the aspect ratios as function of the DDI strength $\epsilon_{\rm dd}$ after release from the trap.  The dashed curve corresponds to the aspect ratio in momentum space $K_{x}/K_{z}$ while the red (light grey) and blue (dark grey) curves correspond to the real-space aspect ratios $R_{y}\lambda_{y}/R_{z}$ and $R_{x}\lambda_{x}/R_{z}$, respectively.}
\label{tof1_asymp}
\end{figure}

A further important quantity of the TOF-analysis of dipolar Fermi gases is the asymptotic values of the aspect ratios. After the expansion the gas becomes more and more dilute and the interaction becomes less and less important, even in the case of long-range ones. Nevertheless, studying the asymptotics in time of the aspect ratios may still be useful because they are approached very fast. This is particularly relevant for strong pancake traps, where this happens just a few $\overline{\omega}^{-1}$ seconds after release of the trap, as is shown in Fig.~\ref{tof1}. Although we are aware of the inaccuracy of the hydrodynamic approach for small dipole moments, we plot the long-time aspect ratios in Fig.~\ref{tof1_asymp} for the whole $\epsilon_{\rm dd}$ range. There, we can identify the tendency of the DDI to stretch the gas in the direction of the applied field in real space, whereas the momentum distribution remains always asymptotically spherical.

We would like to remark that the results presented here are in overall disagreement with those obtained by Sogo et al. in Ref.~\cite{1367-2630-11-5-055017}. Translating their findings into our notation, the ballistic formalism predicts an inversion of the aspect ratio only for $\lambda_{}>1$. In addition, the aspect ratio in real space $R_{x}\lambda_{}/R_{z}$ asymptotically approaches the one in momentum space $K_{x}/K_{z}$ for every value of $\lambda$ and $\epsilon_{\rm dd}$. This disagreement stems from the difference in nature of both approaches: while hydrodynamics assumes local equilibrium provided by collisions, the ballistic approach relies on the assumption of no interaction during the expansion. While the latter might be true for weak interactions, the former seems to be more adequate for strongly interacting gases like the one made out of KRb molecules.

\section{Conclusion \label{conc}}

We have studied both equilibrium and dynamical properties of a normal dipolar Fermi gas in a triaxial harmonic trap. Using a convenient ansatz for the Wigner phase-space function of a normal Fermi gas at very low temperatures, we were able to derive equations of motion which govern the momentum and particle distributions as functions of time, as the trap is shaken or even turned off. The dynamical theory developed here allows, as a special case, to study equilibrium properties starting from the aspect ratios in real and in momentum space and including the stability diagram. Apart from that, the hydrodynamic low-lying excitations were investigated and a level crossing was found in the spectrum, which corresponds to the spherically symmetric limit of a cylinder-symmetric trap. In the case of a triaxial trap with an external field along one of the axes, momentum oscillations were found to be two-dimensional and in-phase, just as for cylinder-symmetric configurations. In addition, we also considered the expansion of the gas after release from the trap, by solving the equations of motion in the absence of the harmonic trap. We found that the characteristic inversion of the aspect ratio in the course of time after release from the trap is also present for dipolar fermions and that the fast experimental development of ultracold heteronuclear KRb-molecules makes them quite promising candidates for observing these effects.

We shall like to remark that the theory presented here fills an important empty space in the study of normal, strong dipolar Fermi gases. Namely, it is applicable in the hydrodynamic regime, where collisions provide local equilibrium. Though the prospects for achieving this regime with heteronuclear molecules are quite exciting, a further gap remains to be filled: Due to the possibility of continuously tuning the interaction strength through an applied electric field, a theory capable of interpolating between the collisionless theory of Ref.~\cite{1367-2630-11-5-055017} and the present hydrodynamic one might be needed in a certain range of the parameter space.

A couple of applications of the present hydrodynamic theory for dipolar Fermi gases could be thought of which would be useful to provide a deeper understanding of these systems. Studying the scissors mode, for example, could shed light on the detection of anisotropic superfluidity. The response of the system to a rotational field $-\Omega {\bf L}$ is a further interesting possibility, which allows to investigate the moment of inertia of a dipolar gas relative to different axes.

\subsection*{Acknowledgements}

We would like to thank J. Dietel, K. Glaum, H. Kleinert, and S. Ospelkaus for useful discussions. We acknowledge financial support from the German Academic Exchange Service (DAAD), from the Innovationsfond of the Freie Universit{\"a}t Berlin, and from the German Research Foundation (DFG) within the Collaborative Research Center SFB/TR12 {\sl Symmetries and Universalities in Mesoscopic Systems}.


\begin{thebibliography}{51}
\expandafter\ifx\csname natexlab\endcsname\relax\def\natexlab#1{#1}\fi
\expandafter\ifx\csname bibnamefont\endcsname\relax
  \def\bibnamefont#1{#1}\fi
\expandafter\ifx\csname bibfnamefont\endcsname\relax
  \def\bibfnamefont#1{#1}\fi
\expandafter\ifx\csname citenamefont\endcsname\relax
  \def\citenamefont#1{#1}\fi
\expandafter\ifx\csname url\endcsname\relax
  \def\url#1{\texttt{#1}}\fi
\expandafter\ifx\csname urlprefix\endcsname\relax\def\urlprefix{URL }\fi
\providecommand{\bibinfo}[2]{#2}
\providecommand{\eprint}[2][]{\url{#2}}

\bibitem[{\citenamefont{Griesmaier et~al.}(2005)\citenamefont{Griesmaier,
  Werner, Hensler, Stuhler, and Pfau}}]{PhysRevLett.94.160401}
\bibinfo{author}{\bibfnamefont{A.}~\bibnamefont{Griesmaier}},
  \bibinfo{author}{\bibfnamefont{J.}~\bibnamefont{Werner}},
  \bibinfo{author}{\bibfnamefont{S.}~\bibnamefont{Hensler}},
  \bibinfo{author}{\bibfnamefont{J.}~\bibnamefont{Stuhler}}, \bibnamefont{and}
  \bibinfo{author}{\bibfnamefont{T.}~\bibnamefont{Pfau}},
  \bibinfo{journal}{Phys. Rev. Lett.} \textbf{\bibinfo{volume}{94}},
  \bibinfo{pages}{160401} (\bibinfo{year}{2005}).

\bibitem[{\citenamefont{Carr and Ye}(2009)}]{1367-2630-11-5-055009}
\bibinfo{author}{\bibfnamefont{L.~D.} \bibnamefont{Carr}} \bibnamefont{and}
  \bibinfo{author}{\bibfnamefont{J.}~\bibnamefont{Ye}}, \bibinfo{journal}{New
  J. Phys.} \textbf{\bibinfo{volume}{11}}, \bibinfo{pages}{055009}
  (\bibinfo{year}{2009}).

\bibitem[{\citenamefont{Lahaye et~al.}(2009)\citenamefont{Lahaye, Menotti,
  Santos, Lewenstein, and Pfau}}]{citeulike:4464283}
\bibinfo{author}{\bibfnamefont{T.}~\bibnamefont{Lahaye}},
  \bibinfo{author}{\bibfnamefont{C.}~\bibnamefont{Menotti}},
  \bibinfo{author}{\bibfnamefont{L.}~\bibnamefont{Santos}},
  \bibinfo{author}{\bibfnamefont{M.}~\bibnamefont{Lewenstein}},
  \bibnamefont{and} \bibinfo{author}{\bibfnamefont{T.}~\bibnamefont{Pfau}},
  \bibinfo{journal}{Rep. Prog. Phys.} \textbf{\bibinfo{volume}{72}},
  \bibinfo{pages}{126401} (\bibinfo{year}{2009}).

\bibitem[{\citenamefont{Ni et~al.}(2008)\citenamefont{Ni, Ospelkaus, {de
  Miranda}, Pe'er, Neyenhuis, Zirbel, Kotochigova, Julienne, Jin, and
  Ye}}]{K.-K.Ni10102008}
\bibinfo{author}{\bibfnamefont{K. K.} \bibnamefont{Ni}},
  \bibinfo{author}{\bibfnamefont{S.}~\bibnamefont{Ospelkaus}},
  \bibinfo{author}{\bibfnamefont{M.~H.~G.} \bibnamefont{{de Miranda}}},
  \bibinfo{author}{\bibfnamefont{A.}~\bibnamefont{Pe'er}},
  \bibinfo{author}{\bibfnamefont{B.}~\bibnamefont{Neyenhuis}},
  \bibinfo{author}{\bibfnamefont{J.~J.} \bibnamefont{Zirbel}},
  \bibinfo{author}{\bibfnamefont{S.}~\bibnamefont{Kotochigova}},
  \bibinfo{author}{\bibfnamefont{P.~S.} \bibnamefont{Julienne}},
  \bibinfo{author}{\bibfnamefont{D.~S.} \bibnamefont{Jin}}, \bibnamefont{and}
  \bibinfo{author}{\bibfnamefont{J.}~\bibnamefont{Ye}},
  \bibinfo{journal}{Science} \textbf{\bibinfo{volume}{322}},
  \bibinfo{pages}{231} (\bibinfo{year}{2008}).

\bibitem[{\citenamefont{Ospelkaus et~al.}(2009)\citenamefont{Ospelkaus, Ni, {de
  Miranda}, Neyenhuis, Wang, Kotochigova, Julienne, Jin, and
  Ye}}]{arXiv:0811.4618}
\bibinfo{author}{\bibfnamefont{S.}~\bibnamefont{Ospelkaus}},
  \bibinfo{author}{\bibfnamefont{K.~K.} \bibnamefont{Ni}},
  \bibinfo{author}{\bibfnamefont{M.~H.~G.} \bibnamefont{{de Miranda}}},
  \bibinfo{author}{\bibfnamefont{B.}~\bibnamefont{Neyenhuis}},
  \bibinfo{author}{\bibfnamefont{D.}~\bibnamefont{Wang}},
  \bibinfo{author}{\bibfnamefont{S.}~\bibnamefont{Kotochigova}},
  \bibinfo{author}{\bibfnamefont{P.~S.} \bibnamefont{Julienne}},
  \bibinfo{author}{\bibfnamefont{D.~S.} \bibnamefont{Jin}}, \bibnamefont{and}
  \bibinfo{author}{\bibfnamefont{J.}~\bibnamefont{Ye}},
  \bibinfo{journal}{Faraday Discuss.} \textbf{\bibinfo{volume}{142}},
  \bibinfo{pages}{351} (\bibinfo{year}{2009}).

\bibitem[{\citenamefont{Ospelkaus et~al.}(2010)\citenamefont{Ospelkaus, Ni,
  Qu{\'e}m{\'e}ner, Neyenhuis, Wang, {de Miranda}, Bohn, Ye, and
  Jin}}]{arXiv:0908.3931}
\bibinfo{author}{\bibfnamefont{S.}~\bibnamefont{Ospelkaus}},
  \bibinfo{author}{\bibfnamefont{K. K.} \bibnamefont{Ni}},
  \bibinfo{author}{\bibfnamefont{G.}~\bibnamefont{Qu{\'e}m{\'e}ner}},
  \bibinfo{author}{\bibfnamefont{B.}~\bibnamefont{Neyenhuis}},
  \bibinfo{author}{\bibfnamefont{D.}~\bibnamefont{Wang}},
  \bibinfo{author}{\bibfnamefont{M.~H.~G.} \bibnamefont{{de Miranda}}},
  \bibinfo{author}{\bibfnamefont{J.~L.} \bibnamefont{Bohn}},
  \bibinfo{author}{\bibfnamefont{J.}~\bibnamefont{Ye}}, \bibnamefont{and}
  \bibinfo{author}{\bibfnamefont{D.~S.} \bibnamefont{Jin}},
  \bibinfo{journal}{Phys. Rev. Lett.} \textbf{\bibinfo{volume}{104}},
  \bibinfo{pages}{030402} (\bibinfo{year}{2010}).

\bibitem[{\citenamefont{Ni et~al.}(2010)\citenamefont{Ni, Ospelkaus, Wang,
  Quemener, Neyenhuis, {de Miranda}, Bohn, Ye, and Jin}}]{citeulike:6565167}
\bibinfo{author}{\bibfnamefont{K. K.} \bibnamefont{Ni}},
  \bibinfo{author}{\bibfnamefont{S.}~\bibnamefont{Ospelkaus}},
  \bibinfo{author}{\bibfnamefont{D.}~\bibnamefont{Wang}},
  \bibinfo{author}{\bibfnamefont{G.}~\bibnamefont{Quemener}},
  \bibinfo{author}{\bibfnamefont{B.}~\bibnamefont{Neyenhuis}},
  \bibinfo{author}{\bibfnamefont{M.~H.~G.} \bibnamefont{{de Miranda}}},
  \bibinfo{author}{\bibfnamefont{J.~L.} \bibnamefont{Bohn}},
  \bibinfo{author}{\bibfnamefont{J.}~\bibnamefont{Ye}}, \bibnamefont{and}
  \bibinfo{author}{\bibfnamefont{D.~S.} \bibnamefont{Jin}},
  \bibinfo{journal}{Nature} \textbf{\bibinfo{volume}{464}},
  \bibinfo{pages}{1324} (\bibinfo{year}{2010}).

\bibitem[{\citenamefont{Yi and You}(2000)}]{PhysRevA.61.041604}
\bibinfo{author}{\bibfnamefont{S.}~\bibnamefont{Yi}} \bibnamefont{and}
  \bibinfo{author}{\bibfnamefont{L.}~\bibnamefont{You}},
  \bibinfo{journal}{Phys. Rev. A} \textbf{\bibinfo{volume}{61}},
  \bibinfo{pages}{041604} (\bibinfo{year}{2000}).

\bibitem[{\citenamefont{O\char39{}Dell
  et~al.}(2004)\citenamefont{O\char39{}Dell, Giovanazzi, and
  Eberlein}}]{PhysRevLett.92.250401}
\bibinfo{author}{\bibfnamefont{D.~H.~J.} \bibnamefont{O\char39{}Dell}},
  \bibinfo{author}{\bibfnamefont{S.}~\bibnamefont{Giovanazzi}},
  \bibnamefont{and} \bibinfo{author}{\bibfnamefont{C.}~\bibnamefont{Eberlein}},
  \bibinfo{journal}{Phys. Rev. Lett.} \textbf{\bibinfo{volume}{92}},
  \bibinfo{pages}{250401} (\bibinfo{year}{2004}).

\bibitem[{\citenamefont{Eberlein et~al.}(2005)\citenamefont{Eberlein,
  Giovanazzi, and O'Dell}}]{eberlein:033618}
\bibinfo{author}{\bibfnamefont{C.}~\bibnamefont{Eberlein}},
  \bibinfo{author}{\bibfnamefont{S.}~\bibnamefont{Giovanazzi}},
  \bibnamefont{and} \bibinfo{author}{\bibfnamefont{D.~H.~J.}
  \bibnamefont{O'Dell}}, \bibinfo{journal}{Phys. Rev. A}
  \textbf{\bibinfo{volume}{71}}, \bibinfo{eid}{033618} (\bibinfo{year}{2005}).

\bibitem[{\citenamefont{Stuhler et~al.}(2005)\citenamefont{Stuhler, Griesmaier,
  Koch, Fattori, Pfau, Giovanazzi, Pedri, and Santos}}]{PhysRevLett.95.150406}
\bibinfo{author}{\bibfnamefont{J.}~\bibnamefont{Stuhler}},
  \bibinfo{author}{\bibfnamefont{A.}~\bibnamefont{Griesmaier}},
  \bibinfo{author}{\bibfnamefont{T.}~\bibnamefont{Koch}},
  \bibinfo{author}{\bibfnamefont{M.}~\bibnamefont{Fattori}},
  \bibinfo{author}{\bibfnamefont{T.}~\bibnamefont{Pfau}},
  \bibinfo{author}{\bibfnamefont{S.}~\bibnamefont{Giovanazzi}},
  \bibinfo{author}{\bibfnamefont{P.}~\bibnamefont{Pedri}}, \bibnamefont{and}
  \bibinfo{author}{\bibfnamefont{L.}~\bibnamefont{Santos}},
  \bibinfo{journal}{Phys. Rev. Lett.} \textbf{\bibinfo{volume}{95}},
  \bibinfo{pages}{150406} (\bibinfo{year}{2005}).

\bibitem[{\citenamefont{Giovanazzi et~al.}(2006)\citenamefont{Giovanazzi,
  Pedri, Santos, Griesmaier, Fattori, Koch, Stuhler, and
  Pfau}}]{giovanazzi:013621}
\bibinfo{author}{\bibfnamefont{S.}~\bibnamefont{Giovanazzi}},
  \bibinfo{author}{\bibfnamefont{P.}~\bibnamefont{Pedri}},
  \bibinfo{author}{\bibfnamefont{L.}~\bibnamefont{Santos}},
  \bibinfo{author}{\bibfnamefont{A.}~\bibnamefont{Griesmaier}},
  \bibinfo{author}{\bibfnamefont{M.}~\bibnamefont{Fattori}},
  \bibinfo{author}{\bibfnamefont{T.}~\bibnamefont{Koch}},
  \bibinfo{author}{\bibfnamefont{J.}~\bibnamefont{Stuhler}}, \bibnamefont{and}
  \bibinfo{author}{\bibfnamefont{T.}~\bibnamefont{Pfau}},
  \bibinfo{journal}{Phys. Rev. A} \textbf{\bibinfo{volume}{74}},
  \bibinfo{eid}{013621} (\bibinfo{year}{2006}).

\bibitem[{\citenamefont{Lahaye et~al.}(2007)\citenamefont{Lahaye, Koch,
  Frohlich, Fattori, Metz, Griesmaier, Giovanazzi, and Pfau}}]{strong-pfau}
\bibinfo{author}{\bibfnamefont{T.}~\bibnamefont{Lahaye}},
  \bibinfo{author}{\bibfnamefont{T.}~\bibnamefont{Koch}},
  \bibinfo{author}{\bibfnamefont{B.}~\bibnamefont{Frohlich}},
  \bibinfo{author}{\bibfnamefont{M.}~\bibnamefont{Fattori}},
  \bibinfo{author}{\bibfnamefont{J.}~\bibnamefont{Metz}},
  \bibinfo{author}{\bibfnamefont{A.}~\bibnamefont{Griesmaier}},
  \bibinfo{author}{\bibfnamefont{S.}~\bibnamefont{Giovanazzi}},
  \bibnamefont{and} \bibinfo{author}{\bibfnamefont{T.}~\bibnamefont{Pfau}},
  \bibinfo{journal}{Nature} \textbf{\bibinfo{volume}{448}},
  \bibinfo{pages}{672} (\bibinfo{year}{2007}).

\bibitem[{\citenamefont{Koch et~al.}(2008)\citenamefont{Koch, Lahaye, Metz,
  Frohlich, Griesmaier, and Pfau}}]{stabilization-pfau}
\bibinfo{author}{\bibfnamefont{T.}~\bibnamefont{Koch}},
  \bibinfo{author}{\bibfnamefont{T.}~\bibnamefont{Lahaye}},
  \bibinfo{author}{\bibfnamefont{J.}~\bibnamefont{Metz}},
  \bibinfo{author}{\bibfnamefont{B.}~\bibnamefont{Frohlich}},
  \bibinfo{author}{\bibfnamefont{A.}~\bibnamefont{Griesmaier}},
  \bibnamefont{and} \bibinfo{author}{\bibfnamefont{T.}~\bibnamefont{Pfau}},
  \bibinfo{journal}{Nature Physics} \textbf{\bibinfo{volume}{4}},
  \bibinfo{pages}{218} (\bibinfo{year}{2008}).

\bibitem[{\citenamefont{Lahaye et~al.}(2008)\citenamefont{Lahaye, Metz,
  Fr{\"o}hlich, Koch, Meister, Griesmaier, Pfau, Saito, Kawaguchi, and
  Ueda}}]{d-wave-pfau}
\bibinfo{author}{\bibfnamefont{T.}~\bibnamefont{Lahaye}},
  \bibinfo{author}{\bibfnamefont{J.}~\bibnamefont{Metz}},
  \bibinfo{author}{\bibfnamefont{B.}~\bibnamefont{Fr{\"o}hlich}},
  \bibinfo{author}{\bibfnamefont{T.}~\bibnamefont{Koch}},
  \bibinfo{author}{\bibfnamefont{M.}~\bibnamefont{Meister}},
  \bibinfo{author}{\bibfnamefont{A.}~\bibnamefont{Griesmaier}},
  \bibinfo{author}{\bibfnamefont{T.}~\bibnamefont{Pfau}},
  \bibinfo{author}{\bibfnamefont{H.}~\bibnamefont{Saito}},
  \bibinfo{author}{\bibfnamefont{Y.}~\bibnamefont{Kawaguchi}},
  \bibnamefont{and} \bibinfo{author}{\bibfnamefont{M.}~\bibnamefont{Ueda}},
  \bibinfo{journal}{Phys. Rev. Lett.} \textbf{\bibinfo{volume}{101}},
  \bibinfo{eid}{080401} (\bibinfo{year}{2008}).

\bibitem[{\citenamefont{Glaum et~al.}(2007)\citenamefont{Glaum, Pelster,
  Kleinert, and Pfau}}]{glaum:080407}
\bibinfo{author}{\bibfnamefont{K.}~\bibnamefont{Glaum}},
  \bibinfo{author}{\bibfnamefont{A.}~\bibnamefont{Pelster}},
  \bibinfo{author}{\bibfnamefont{H.}~\bibnamefont{Kleinert}}, \bibnamefont{and}
  \bibinfo{author}{\bibfnamefont{T.}~\bibnamefont{Pfau}},
  \bibinfo{journal}{Phys. Rev. Lett.} \textbf{\bibinfo{volume}{98}},
  \bibinfo{eid}{080407} (\bibinfo{year}{2007}).

\bibitem[{\citenamefont{Glaum and Pelster}(2007)}]{glaum:023604}
\bibinfo{author}{\bibfnamefont{K.}~\bibnamefont{Glaum}} \bibnamefont{and}
  \bibinfo{author}{\bibfnamefont{A.}~\bibnamefont{Pelster}},
  \bibinfo{journal}{Phys. Rev. A} \textbf{\bibinfo{volume}{76}},
  \bibinfo{eid}{023604} (\bibinfo{year}{2007}).

\bibitem[{\citenamefont{Kawaguchi et~al.}(2006)\citenamefont{Kawaguchi, Saito,
  and Ueda}}]{kawaguchi:080405}
\bibinfo{author}{\bibfnamefont{Y.}~\bibnamefont{Kawaguchi}},
  \bibinfo{author}{\bibfnamefont{H.}~\bibnamefont{Saito}}, \bibnamefont{and}
  \bibinfo{author}{\bibfnamefont{M.}~\bibnamefont{Ueda}},
  \bibinfo{journal}{Phys. Rev. Lett.} \textbf{\bibinfo{volume}{96}},
  \bibinfo{eid}{080405} (\bibinfo{year}{2006}).

\bibitem[{\citenamefont{Ronen and Bohn}(2009)}]{citeulike:6646198}
\bibinfo{author}{\bibfnamefont{S.}~\bibnamefont{Ronen}} \bibnamefont{and}
  \bibinfo{author}{\bibfnamefont{J.~L.} \bibnamefont{Bohn}},
   \bibinfo{journal}{Phys. Rev. A} \textbf{\bibinfo{volume}{81}},
  \bibinfo{pages}{033601} (\bibinfo{year}{2010}).

\bibitem[{\citenamefont{Chan et~al.}(2010)\citenamefont{Chan, Wu, Lee, and
  Sarma}}]{citeulike:5114415}
\bibinfo{author}{\bibfnamefont{C.-K.} \bibnamefont{Chan}},
  \bibinfo{author}{\bibfnamefont{C.}~\bibnamefont{Wu}},
  \bibinfo{author}{\bibfnamefont{W.-C.} \bibnamefont{Lee}}, \bibnamefont{and}
  \bibinfo{author}{\bibfnamefont{S.~D.} \bibnamefont{Sarma}},
  \bibinfo{journal}{Phys. Rev. A} \textbf{\bibinfo{volume}{81}},
  \bibinfo{pages}{023602} (\bibinfo{year}{2010}).

\bibitem[{\citenamefont{Bruun and Taylor}(2008)}]{bruun:245301}
\bibinfo{author}{\bibfnamefont{G.~M.} \bibnamefont{Bruun}} \bibnamefont{and}
  \bibinfo{author}{\bibfnamefont{E.}~\bibnamefont{Taylor}},
  \bibinfo{journal}{Phys. Rev. Lett.} \textbf{\bibinfo{volume}{101}},
  \bibinfo{eid}{245301} (\bibinfo{year}{2008}).

\bibitem[{\citenamefont{Fregoso et~al.}(2009)\citenamefont{Fregoso, Sun,
  Fradkin, and Lev}}]{1367-2630-11-10-103003}
\bibinfo{author}{\bibfnamefont{B.~M.} \bibnamefont{Fregoso}},
  \bibinfo{author}{\bibfnamefont{K.}~\bibnamefont{Sun}},
  \bibinfo{author}{\bibfnamefont{E.}~\bibnamefont{Fradkin}}, \bibnamefont{and}
  \bibinfo{author}{\bibfnamefont{B.~L.} \bibnamefont{Lev}},
  \bibinfo{journal}{New J. Phys.} \textbf{\bibinfo{volume}{11}},
  \bibinfo{pages}{103003} (\bibinfo{year}{2009}).

\bibitem[{\citenamefont{Fregoso and Fradkin}(2009)}]{fregoso:205301}
\bibinfo{author}{\bibfnamefont{B.~M.} \bibnamefont{Fregoso}} \bibnamefont{and}
  \bibinfo{author}{\bibfnamefont{E.}~\bibnamefont{Fradkin}},
  \bibinfo{journal}{Phys. Rev. Lett.} \textbf{\bibinfo{volume}{103}},
  \bibinfo{eid}{205301} (\bibinfo{year}{2009}).

\bibitem[{\citenamefont{Baranov et~al.}(2004)\citenamefont{Baranov, Dobrek, and
  Lewenstein}}]{baranov:250403}
\bibinfo{author}{\bibfnamefont{M.~A.} \bibnamefont{Baranov}},
  \bibinfo{author}{\bibfnamefont{{\L}.}~\bibnamefont{Dobrek}},
  \bibnamefont{and}
  \bibinfo{author}{\bibfnamefont{M.}~\bibnamefont{Lewenstein}},
  \bibinfo{journal}{Phys. Rev. Lett.} \textbf{\bibinfo{volume}{92}},
  \bibinfo{eid}{250403} (\bibinfo{year}{2004}).

\bibitem[{\citenamefont{Baranov et~al.}(2005)\citenamefont{Baranov, Osterloh,
  and Lewenstein}}]{baranov:070404}
\bibinfo{author}{\bibfnamefont{M.~A.} \bibnamefont{Baranov}},
  \bibinfo{author}{\bibfnamefont{K.}~\bibnamefont{Osterloh}}, \bibnamefont{and}
  \bibinfo{author}{\bibfnamefont{M.}~\bibnamefont{Lewenstein}},
  \bibinfo{journal}{Phys. Rev. Lett.} \textbf{\bibinfo{volume}{94}},
  \bibinfo{eid}{070404} (\bibinfo{year}{2005}).

\bibitem[{\citenamefont{Baranov et~al.}(2008)\citenamefont{Baranov, Fehrmann,
  and Lewenstein}}]{baranov:200402}
\bibinfo{author}{\bibfnamefont{M.~A.} \bibnamefont{Baranov}},
  \bibinfo{author}{\bibfnamefont{H.}~\bibnamefont{Fehrmann}}, \bibnamefont{and}
  \bibinfo{author}{\bibfnamefont{M.}~\bibnamefont{Lewenstein}},
  \bibinfo{journal}{Phys. Rev. Lett.} \textbf{\bibinfo{volume}{100}},
  \bibinfo{eid}{200402} (\bibinfo{year}{2008}).

\bibitem[{\citenamefont{Chicireanu et~al.}(2006)\citenamefont{Chicireanu,
  Pouderous, Barb{\'e}, Laburthe-Tolra, Mar{\'e}chal, Vernac, Keller, and
  Gorceix}}]{chicireanu:053406}
\bibinfo{author}{\bibfnamefont{R.}~\bibnamefont{Chicireanu}},
  \bibinfo{author}{\bibfnamefont{A.}~\bibnamefont{Pouderous}},
  \bibinfo{author}{\bibfnamefont{R.}~\bibnamefont{Barb{\'e}}},
  \bibinfo{author}{\bibfnamefont{B.}~\bibnamefont{Laburthe-Tolra}},
  \bibinfo{author}{\bibfnamefont{E.}~\bibnamefont{Mar{\'e}chal}},
  \bibinfo{author}{\bibfnamefont{L.}~\bibnamefont{Vernac}},
  \bibinfo{author}{\bibfnamefont{J.-C.} \bibnamefont{Keller}},
  \bibnamefont{and} \bibinfo{author}{\bibfnamefont{O.}~\bibnamefont{Gorceix}},
  \bibinfo{journal}{Phys. Rev. A} \textbf{\bibinfo{volume}{73}},
  \bibinfo{eid}{053406} (\bibinfo{year}{2006}).

\bibitem[{\citenamefont{Fukuhara et~al.}(2007)\citenamefont{Fukuhara, Takasu,
  Kumakura, and Takahashi}}]{fukuhara:030401}
\bibinfo{author}{\bibfnamefont{T.}~\bibnamefont{Fukuhara}},
  \bibinfo{author}{\bibfnamefont{Y.}~\bibnamefont{Takasu}},
  \bibinfo{author}{\bibfnamefont{M.}~\bibnamefont{Kumakura}}, \bibnamefont{and}
  \bibinfo{author}{\bibfnamefont{Y.}~\bibnamefont{Takahashi}},
  \bibinfo{journal}{Phys. Rev. Lett.} \textbf{\bibinfo{volume}{98}},
  \bibinfo{eid}{030401} (\bibinfo{year}{2007}).

\bibitem[{\citenamefont{Lu et~al.}(2010)\citenamefont{Lu, Youn, and
  Lev}}]{dysp_experiment_two}
\bibinfo{author}{\bibfnamefont{M.}~\bibnamefont{Lu}},
  \bibinfo{author}{\bibfnamefont{S.~H.} \bibnamefont{Youn}}, \bibnamefont{and}
  \bibinfo{author}{\bibfnamefont{B.~L.} \bibnamefont{Lev}},
  \bibinfo{journal}{Phys. Rev. Lett.} \textbf{\bibinfo{volume}{104}},
  \bibinfo{pages}{063001} (\bibinfo{year}{2010}).

\bibitem[{\citenamefont{Ticknor}(2008)}]{ticknor:133202}
\bibinfo{author}{\bibfnamefont{C.}~\bibnamefont{Ticknor}},
  \bibinfo{journal}{Phys. Rev. Lett.} \textbf{\bibinfo{volume}{100}},
  \bibinfo{eid}{133202} (\bibinfo{year}{2008}).

\bibitem[{\citenamefont{Bohn et~al.}(2009)\citenamefont{Bohn, Cavagnero, and
  Ticknor}}]{1367-2630-11-5-055039}
\bibinfo{author}{\bibfnamefont{J.~L.} \bibnamefont{Bohn}},
  \bibinfo{author}{\bibfnamefont{M.}~\bibnamefont{Cavagnero}},
  \bibnamefont{and} \bibinfo{author}{\bibfnamefont{C.}~\bibnamefont{Ticknor}},
  \bibinfo{journal}{New J. Phys.} \textbf{\bibinfo{volume}{11}},
  \bibinfo{pages}{055039} (\bibinfo{year}{2009}).

\bibitem[{\citenamefont{Bourdel et~al.}(2003)\citenamefont{Bourdel, Cubizolles,
  Khaykovich, Magalhaes, Kokkelmans, Shlyapnikov, and
  Salomon}}]{PhysRevLett.91.020402}
\bibinfo{author}{\bibfnamefont{T.}~\bibnamefont{Bourdel}},
  \bibinfo{author}{\bibfnamefont{J.}~\bibnamefont{Cubizolles}},
  \bibinfo{author}{\bibfnamefont{L.}~\bibnamefont{Khaykovich}},
  \bibinfo{author}{\bibfnamefont{K.~M.~F.} \bibnamefont{Magalhaes}},
  \bibinfo{author}{\bibfnamefont{S.~J. J. M.~F.} \bibnamefont{Kokkelmans}},
  \bibinfo{author}{\bibfnamefont{G.~V.} \bibnamefont{Shlyapnikov}},
  \bibnamefont{and} \bibinfo{author}{\bibfnamefont{C.}~\bibnamefont{Salomon}},
  \bibinfo{journal}{Phys. Rev. Lett.} \textbf{\bibinfo{volume}{91}},
  \bibinfo{pages}{020402} (\bibinfo{year}{2003}).

\bibitem[{\citenamefont{G{\'o}ral et~al.}(2001)\citenamefont{G{\'o}ral,
  Englert, and {Rza\ifmmode \mbox{\c{}}\else \c{}\fi{}\ifmmode \dot{z}\else
  \.{z}\fi{}ewski}}}]{PhysRevA.63.033606}
\bibinfo{author}{\bibfnamefont{K.}~\bibnamefont{G{\'o}ral}},
  \bibinfo{author}{\bibfnamefont{B.-G.} \bibnamefont{Englert}},
  \bibnamefont{and}
  \bibinfo{author}{\bibfnamefont{K.}~\bibnamefont{{Rza\ifmmode \mbox{\c{}}\else
  \c{}\fi{}\ifmmode \dot{z}\else \.{z}\fi{}ewski}}}, \bibinfo{journal}{Phys.
  Rev. A} \textbf{\bibinfo{volume}{63}}, \bibinfo{pages}{033606}
  (\bibinfo{year}{2001}).

\bibitem[{\citenamefont{G{\'o}ral et~al.}(2003)\citenamefont{G{\'o}ral,
  Brewczyk, and {Rza\ifmmode \mbox{\c{}}\else \c{}\fi{}\ifmmode \dot{z}\else
  \.{z}\fi{}ewski}}}]{PhysRevA.67.025601}
\bibinfo{author}{\bibfnamefont{K.}~\bibnamefont{G{\'o}ral}},
  \bibinfo{author}{\bibfnamefont{M.}~\bibnamefont{Brewczyk}}, \bibnamefont{and}
  \bibinfo{author}{\bibfnamefont{K.}~\bibnamefont{{Rza\ifmmode \mbox{\c{}}\else
  \c{}\fi{}\ifmmode \dot{z}\else \.{z}\fi{}ewski}}}, \bibinfo{journal}{Phys.
  Rev. A} \textbf{\bibinfo{volume}{67}}, \bibinfo{pages}{025601}
  (\bibinfo{year}{2003}).

\bibitem[{\citenamefont{Miyakawa et~al.}(2008)\citenamefont{Miyakawa, Sogo, and
  Pu}}]{miyakawa:061603}
\bibinfo{author}{\bibfnamefont{T.}~\bibnamefont{Miyakawa}},
  \bibinfo{author}{\bibfnamefont{T.}~\bibnamefont{Sogo}}, \bibnamefont{and}
  \bibinfo{author}{\bibfnamefont{H.}~\bibnamefont{Pu}}, \bibinfo{journal}{Phys.
  Rev. A} \textbf{\bibinfo{volume}{77}}, \bibinfo{eid}{061603(R)}
  (\bibinfo{year}{2008}).

\bibitem[{\citenamefont{Sogo et~al.}(2009)\citenamefont{Sogo, He, Miyakawa, Yi,
  Lu, and Pu}}]{1367-2630-11-5-055017}
\bibinfo{author}{\bibfnamefont{T.}~\bibnamefont{Sogo}},
  \bibinfo{author}{\bibfnamefont{L.}~\bibnamefont{He}},
  \bibinfo{author}{\bibfnamefont{T.}~\bibnamefont{Miyakawa}},
  \bibinfo{author}{\bibfnamefont{S.}~\bibnamefont{Yi}},
  \bibinfo{author}{\bibfnamefont{H.}~\bibnamefont{Lu}}, \bibnamefont{and}
  \bibinfo{author}{\bibfnamefont{H.}~\bibnamefont{Pu}}, \bibinfo{journal}{New
  J. Phys.} \textbf{\bibinfo{volume}{11}}, \bibinfo{pages}{055017}
  (\bibinfo{year}{2009}).

\bibitem[{\citenamefont{He et~al.}(2008)\citenamefont{He, Zhang, Zhang, and
  Yi}}]{he:031605}
\bibinfo{author}{\bibfnamefont{L.}~\bibnamefont{He}},
  \bibinfo{author}{\bibfnamefont{J.-N.} \bibnamefont{Zhang}},
  \bibinfo{author}{\bibfnamefont{Y.}~\bibnamefont{Zhang}}, \bibnamefont{and}
  \bibinfo{author}{\bibfnamefont{S.}~\bibnamefont{Yi}}, \bibinfo{journal}{Phys.
  Rev. A} \textbf{\bibinfo{volume}{77}}, \bibinfo{eid}{031605}
  (\bibinfo{year}{2008}).

\bibitem[{\citenamefont{Lima and Pelster}(2010)}]{ourpaper}
\bibinfo{author}{\bibfnamefont{A.~R.~P.} \bibnamefont{Lima}} \bibnamefont{and}
  \bibinfo{author}{\bibfnamefont{A.}~\bibnamefont{Pelster}},
  \bibinfo{journal}{Phys. Rev. A} \textbf{\bibinfo{volume}{81}},
  \bibinfo{pages}{021606(R)} (\bibinfo{year}{2010}).

\bibitem[{\citenamefont{Giannoni et~al.}(1976)\citenamefont{Giannoni,
  Vautherin, Veneroni, and Brink}}]{brink}
\bibinfo{author}{\bibfnamefont{M.~J.} \bibnamefont{Giannoni}},
  \bibinfo{author}{\bibfnamefont{D.}~\bibnamefont{Vautherin}},
  \bibinfo{author}{\bibfnamefont{M.}~\bibnamefont{Veneroni}}, \bibnamefont{and}
  \bibinfo{author}{\bibfnamefont{D.~M.} \bibnamefont{Brink}},
  \bibinfo{journal}{Phys. Lett. B} \textbf{\bibinfo{volume}{63}},
  \bibinfo{pages}{8} (\bibinfo{year}{1976}).

\bibitem[{\citenamefont{Lipparini}(2003)}]{lipparini}
\bibinfo{author}{\bibfnamefont{E.}~\bibnamefont{Lipparini}},
  \emph{\bibinfo{title}{Modern Many-Particle Physics, {\sl Atomic Gases,
  Quantum Dots and Quantum Fluids}}} (\bibinfo{publisher}{World Scientific},
  \bibinfo{year}{2003}).

\bibitem[{\citenamefont{Bruun and Clark}(1999)}]{PhysRevLett.83.5415}
\bibinfo{author}{\bibfnamefont{G.~M.} \bibnamefont{Bruun}} \bibnamefont{and}
  \bibinfo{author}{\bibfnamefont{C.~W.} \bibnamefont{Clark}},
  \bibinfo{journal}{Phys. Rev. Lett.} \textbf{\bibinfo{volume}{83}},
  \bibinfo{pages}{5415} (\bibinfo{year}{1999}).

\bibitem[{\citenamefont{Amoruso et~al.}(1999)\citenamefont{Amoruso, Meccoli,
  Minguzzi, and Tosi}}]{amoruso}
\bibinfo{author}{\bibfnamefont{M.}~\bibnamefont{Amoruso}},
  \bibinfo{author}{\bibfnamefont{I.}~\bibnamefont{Meccoli}},
  \bibinfo{author}{\bibfnamefont{A.}~\bibnamefont{Minguzzi}}, \bibnamefont{and}
  \bibinfo{author}{\bibfnamefont{M.}~\bibnamefont{Tosi}},
  \bibinfo{journal}{Eur. Phys. J. D} \textbf{\bibinfo{volume}{7}},
  \bibinfo{pages}{441} (\bibinfo{year}{1999}).

\bibitem[{\citenamefont{Ring and Schuck}(2000)}]{ring-schuck}
\bibinfo{author}{\bibfnamefont{P.}~\bibnamefont{Ring}} \bibnamefont{and}
  \bibinfo{author}{\bibfnamefont{P.}~\bibnamefont{Schuck}},
  \emph{\bibinfo{title}{The Nuclear Many-Body Problem}}
  (\bibinfo{publisher}{Springer}, \bibinfo{year}{2000}).

\bibitem[{\citenamefont{Hohenberg and Kohn}(1964)}]{PhysRev.136.B864}
\bibinfo{author}{\bibfnamefont{P.}~\bibnamefont{Hohenberg}} \bibnamefont{and}
  \bibinfo{author}{\bibfnamefont{W.}~\bibnamefont{Kohn}},
  \bibinfo{journal}{Phys. Rev.} \textbf{\bibinfo{volume}{136}},
  \bibinfo{pages}{B864} (\bibinfo{year}{1964}).

\bibitem[{\citenamefont{Pethick and Smith}(2008)}]{pethick}
\bibinfo{author}{\bibfnamefont{C.~J.} \bibnamefont{Pethick}} \bibnamefont{and}
  \bibinfo{author}{\bibfnamefont{H.}~\bibnamefont{Smith}},
  \emph{\bibinfo{title}{Bose-Einstein Condensation in Dilute Gases}},
  \bibinfo{edition}{2nd} ed
  (\bibinfo{publisher}{Cambridge University Press}, \bibinfo{year}{2008}).

\bibitem[{\citenamefont{Yi and You}(2001)}]{PhysRevA.63.053607}
\bibinfo{author}{\bibfnamefont{S.}~\bibnamefont{Yi}} \bibnamefont{and}
  \bibinfo{author}{\bibfnamefont{L.}~\bibnamefont{You}},
  \bibinfo{journal}{Phys. Rev. A} \textbf{\bibinfo{volume}{63}},
  \bibinfo{pages}{053607} (\bibinfo{year}{2001}).

\bibitem[{\citenamefont{Csord{\'a}s and Graham}(2000)}]{PhysRevA.63.013606}
\bibinfo{author}{\bibfnamefont{A.}~\bibnamefont{Csord{\'a}s}} \bibnamefont{and}
  \bibinfo{author}{\bibfnamefont{R.}~\bibnamefont{Graham}},
  \bibinfo{journal}{Phys. Rev. A} \textbf{\bibinfo{volume}{63}},
  \bibinfo{pages}{013606} (\bibinfo{year}{2000}).

\bibitem[{\citenamefont{Fliesser et~al.}(1997)\citenamefont{Fliesser,
  Csord{\'a}s, Sz{\'e}pfalusy, and Graham}}]{PhysRevA.56.R2533}
\bibinfo{author}{\bibfnamefont{M.}~\bibnamefont{Fliesser}},
  \bibinfo{author}{\bibfnamefont{A.}~\bibnamefont{Csord{\'a}s}},
  \bibinfo{author}{\bibfnamefont{P.}~\bibnamefont{Sz{\'e}pfalusy}},
  \bibnamefont{and} \bibinfo{author}{\bibfnamefont{R.}~\bibnamefont{Graham}},
  \bibinfo{journal}{Phys. Rev. A} \textbf{\bibinfo{volume}{56}},
  \bibinfo{pages}{2533(R)} (\bibinfo{year}{1997}).

\bibitem[{\citenamefont{Altmeyer et~al.}(2007)\citenamefont{Altmeyer, Riedl,
  Wright, Kohstall, Denschlag, and Grimm}}]{altmeyer:033610}
\bibinfo{author}{\bibfnamefont{A.}~\bibnamefont{Altmeyer}},
  \bibinfo{author}{\bibfnamefont{S.}~\bibnamefont{Riedl}},
  \bibinfo{author}{\bibfnamefont{M.~J.} \bibnamefont{Wright}},
  \bibinfo{author}{\bibfnamefont{C.}~\bibnamefont{Kohstall}},
  \bibinfo{author}{\bibfnamefont{J.~H.} \bibnamefont{Denschlag}},
  \bibnamefont{and} \bibinfo{author}{\bibfnamefont{R.}~\bibnamefont{Grimm}},
  \bibinfo{journal}{Phys. Rev. A} \textbf{\bibinfo{volume}{76}},
  \bibinfo{eid}{033610} (\bibinfo{year}{2007}).

\bibitem[{\citenamefont{Giovanazzi et~al.}(2007)\citenamefont{Giovanazzi,
  Santos, and Pfau}}]{PhysRevA.75.015604}
\bibinfo{author}{\bibfnamefont{S.}~\bibnamefont{Giovanazzi}},
  \bibinfo{author}{\bibfnamefont{L.}~\bibnamefont{Santos}}, \bibnamefont{and}
  \bibinfo{author}{\bibfnamefont{T.}~\bibnamefont{Pfau}},
  \bibinfo{journal}{Phys. Rev. A} \textbf{\bibinfo{volume}{75}},
  \bibinfo{pages}{015604} (\bibinfo{year}{2007}).

\bibitem[{\citenamefont{O'Hara et~al.}(2002)\citenamefont{O'Hara, Hemmer, Gehm,
  Granade, and Thomas}}]{ohara}
\bibinfo{author}{\bibfnamefont{K.~M.} \bibnamefont{O'Hara}},
  \bibinfo{author}{\bibfnamefont{S.~L.} \bibnamefont{Hemmer}},
  \bibinfo{author}{\bibfnamefont{M.~E.} \bibnamefont{Gehm}},
  \bibinfo{author}{\bibfnamefont{S.~R.} \bibnamefont{Granade}},
  \bibnamefont{and} \bibinfo{author}{\bibfnamefont{J.~E.}
  \bibnamefont{Thomas}}, \bibinfo{journal}{Science}
  \textbf{\bibinfo{volume}{298}}, \bibinfo{pages}{2179} (\bibinfo{year}{2002}).

\end{thebibliography}
\end{document}